\documentclass[prb,twocolumn]{revtex4-1}
\usepackage{amsmath,amssymb,mathrsfs}
\usepackage{psfrag}
\usepackage{appendix}
\usepackage{graphicx}
\usepackage{graphics}
\usepackage{epsfig}
\usepackage{bm}
\usepackage{color}
\usepackage{verbatim,color,ulem}

\usepackage{tikz}
\usetikzlibrary{calc,decorations.markings}
\usetikzlibrary{decorations.markings}
\tikzset
  {decoration=
     {markings,
      mark=at position 0.5 with {\arrow{stealth}}
     },
   plain/.style={line width=0.8pt},
   arrow/.style={plain,postaction=decorate}
  }
\usepackage{mathtools}
\usetikzlibrary{external}
\usetikzlibrary{shapes.misc}

\tikzset{cross/.style={cross out, draw=black, minimum size=2*(#1-\pgflinewidth), inner sep=0pt, outer sep=0pt},   cross/.default={1pt}}

\usetikzlibrary{decorations.pathmorphing}

\newcommand{\beq}{\begin{equation}}
\newcommand{\eeq}{\end{equation}}
\newcommand{\beqa}{\begin{eqnarray}}
\newcommand{\eeqa}{\end{eqnarray}}
\newcommand{\ba}{\begin{aligned}[b]}
\newcommand{\ea}{\end{aligned}}
\newcommand{\mO}{\mathcal{O}}
\newcommand{\mZ}{\mathcal{Z}}

\newcommand{\no}{\text{no}}
\newcommand{\co}{\text{co}}

\newcommand{\hs}{\hat{s}}
\newcommand{\cpa}{\textrm{CP}_1(\varrho)}
\newcommand{\cpb}{\textrm{CP}_2(\varrho)}
\newcommand{\cpc}{\textrm{CP}_3(\varrho)}
\newcommand{\gcol}[1]{{\color{black} #1}}

\begin{document}


\title{Ideal Bose-Einstein condensation in the canonical
  ensemble:\\ exact asymptotic estimates from large deviations}

\author{Giacomo Gradenigo} \affiliation{Dipartimento di Fisica e
  Astronomia ``Galileo Galilei'', Universit\`a di Padova, via Marzolo
  8, I-35131 Padova, Italy} \affiliation{Istituto Nazionale di Fisica
  Nucleare (INFN), Sezione di Padova, I-35131 Padova, Italy}
\email{giacomo.gradenigo@unipd.it}

\author{Dario Lucente} \affiliation{Dipartimento di Fisica,
  Universit\`a di Roma ``La Sapienza'', P.le Aldo Moro 5, 00185, Rome,
  Italy}

\author{Luca Salasnich} \affiliation{Dipartimento di Fisica e
  Astronomia ``Galileo Galilei'' and QTech, Universit\`a di Padova,
  via Marzolo 8, I-35131 Padova, Italy} \affiliation{Istituto
  Nazionale di Fisica Nucleare (INFN), Sezione di Padova, I-35131
  Padova, Italy} 
\email{luca.salasnich@unipd.it}

\date{\today}

\begin{abstract}
In this work we present a large-deviations approach to the calculation
of the canonical partition function for free bosons. Three-dimensional
Bose-Einstein condensation is studied in the fixed-density ensemble as
a function of the dimensionless density $\varrho = \rho \lambda_T^3$,
with $\rho=N/L^3$ the standard particle density, $\lambda_T$ the
thermal wavelength, $L$ the linear size of the box and $N$ the total
number of particles. A large-deviations approach in terms of the
dimensionless parameter $\ell=L/\lambda_T$ allows us to provide exact
asymptotic estimates of the canonical partition function both above
and below the critical density $\varrho_c$ for Bose-Einstein
condensation. We show how this approach allows to explicitly account
for finite-size effects and how it fully captures the first-order
aspects of the transition, allowing us to explicitate its driving
mechanism in terms of the competing probabilities of normal and
condensed phases. The proposed large-deviations approach allows then
to obtain in all regimes explicit and simple analytical expressions,
at the leading order in the large parameter $\ell$, for both the
average fraction of particles in the ground state, the condensate
fraction $\langle n_0(\varrho) \rangle = \langle N_0(\varrho)
\rangle/N$, and for its fluctuations, $\sigma_0(\varrho) =
\sqrt{\langle N_0^2(\varrho)\rangle - \langle
  N_0(\varrho)\rangle^2}/N$, retrieving for instance the anomalous
scaling $\sigma_0(\varrho)\sim 1/V^{1/3}$ in the condensed regime,
$\varrho > \varrho_c$. Our large-deviations asymptotic estimate, by
analytically clarifying the mixed-order nature of Bose-Einstein
condensation, allows then to reveal the similarity between this
transition and other mixed-order transitions, as for instance the
localization transition in the Discrete Non-Linear Schr\"odinger
Equation.
\end{abstract}

\maketitle
\tableofcontents

\section{Introduction}
\label{sec:intro}

It is a quite remarkable fact that the nature of Bose–Einstein
Condensation (BEC) in an ideal (non-interacting) system of bosons
remains even nowadays elusive. Unlike interacting systems, where BEC
is firmly established as a second-order phase
transition~\cite{L01,SP16}, in the ideal case, in which the transition
is driven solely by a global constraint on particle number, the order
of the transition cannot be easily ascertained. The most extended and
detailed technical account of ideal Bose-Einstein Condensation can be
found in the work of Ziff, Kac, and Uhlenbeck~\cite{ZUK77}, in which
also another subtle point of Bose-Einstein condensation is thoroughly
accounted: anomalies for the predictions of the grand-canonical
ensemble in the condensed phase and their discrepancy with that of the
fixed density ensemble, as for instance the grand-canonical divergence
of condensate fraction fluctuations. This point is particularly
interesting because recent experimental observations of
grand-canonical statistics in photon condensates~\cite{SDDVKW14},
along with new theoretical developments~\cite{CSZ26}, have reopened
the debate on the ensemble-dependent predictions for BEC. Beside the
ensemble-dependent analytical predictions on the condensed phase
behaviour, another intriguing aspect of BEC is that, as we just said,
ideal Bose-Einstein condensation exhibits the features of a
mixed-order (first/second) transition: on the one hand it shows
properties, like phase coexistence, which are typical of first-order
transitions, a perspective which has been particularly emphasized by
an ``historical'' reference such as the statistical mechanics book of
Kerson Huang~\cite{H87}; on the other hand ideal Bose-Einstein
condensation also displays features of continuous transitions, like
the absence of specific heat or the presence of symmetry
breaking~\cite{SP16,CSZ26}, which motivate people, in particular
within the condensed-matter community~\cite{SP16}, to classify it as a
second-order phase transition. The canonical approach to Bose-Einstein
condensation discussed here reveals its similarity with other
mixed-order transitions, as for instance the localization transition
in the Discrete Nonlinear Schr\"odinger Equation
(DNLSE)~\cite{RCKG00,GILM21,GILM21b} or even some sort of ideal glass
transitions, as for instance the one found in the Random Energy
Model~\cite{P24}.\\

We show here how the long-standing conundrum of the mixed-order
features of Bose-Einstein Condensation for free bosons can be resolved
by applying a large-deviation strategy to the calculation of the
canonical partition function, extending to BEC the analytical approach
which furnished exact results in the localized regime of the DNLSE, in
which the localized phase is characterized by physical features deeply
connected to the lack of equivalence between statistical
ensembles~\cite{GILM21,GILM21b,MEZ05,EMZ06,SEM14,GB17,MGM21}. On the
one hand, our approach allows to highlight the first-order features of
the transition; on the other hand, it is only the canonical analysis
of Bose-Einstein condensation which allows to study the average value
and the fluctuations of the condensate fraction for (dimensionless)
densities both below, $\varrho<\varrho_c$, and above,
$\varrho>\varrho_c$, the critical one $\varrho_c$. Historically, a
rigorous and comprehensive calculation of the canonical partition
function for free bosons can be found in the work of Ziff {\it et
  al.}~\cite{ZUK77}, together with the statement that {\it ``in the
  thermodynamic limit''} the grand-canonical ensemble fails to
describe the condensed phase properties. Let us recall, to this
concern, that formally the inequivalence of statistical ensembles,
both in the form of different quantitative predictions for the two
ensembles (as is, for instance, the usual case for long-range
forces~\cite{CDR09}) or in the form of the complete failure of one of
the two in describing a certain phase, it is something well defined
only in the thermodynamic limit. The goal of this proposal is to
provide simple and explicit expressions for the average condensate
fraction and for its fluctuations in the BEC regime, drawn from the
fixed density ensemble and which remain well-defined in the
thermodynamic limit. While showing that this degree of accuracy and
consistency can only be achieved in the canonical ensemble, we by no
mean question the validity of grand-canonical results, {\it as long as
  finite system sizes are considered}, a situation which can be for
instance reproduced in ad-hoc experimental setups where BEC is
realized with photons~\cite{SDDVKW14} rather than cold atoms. Our
results represents a real technical improvement with respect to the
results of~\cite{ZUK77}, since our large-deviation approach allows to
simplify the calculations and to analytically obtain simple and
explicit expression, in particular for the system-size dependence of
the average condensate fraction $\langle n_0(\varrho) \rangle =
\langle N_0(\varrho) \rangle/N$ and its fluctuations
$\sigma_0(\varrho) \sim \sqrt{\langle N_0^2(\varrho) \rangle - \langle
  N_0(\varrho) \rangle^2}/N$. The first precise calculation of
fluctuations statistics in ensembles where the number of particles was
conserved dates back to Politzer~\cite{P96}, with recent experimental
confirmation from the study of cold atoms
condensates~\cite{CGIPKSRHA19,CVHKPHRKA21}. Nevertheless, such
theoretical estimates are provided by non explicit numerical sampling
of the canonical partition function~\cite{WW97}, as thoroughly
discussed in a recent very comprehensive review of fluctuations in
cold atoms systems~\cite{KKADGPWAR25}. Our method is the proposal of a
robust framework to obtain explicitly asymptotic estimates in the
canonical ensemble, with the possibility to be extended, in future
works, to the microcanonical one. About the mixed order of the
transition, our approach allows to clearly show how the
transition-mechanism, despite the absence of latent heat which usually
accompanies phase coexistence, still is controlled by the competition
between two phases, to which the canonical approach allows to assign a
well-defined probability. As a last preliminary comment, let us notice
that the analysis presented here has a purpose similar to the refined
saddle-point treatment of condensed Bose gases by Holthaus and
Kalinowski~\cite{HK99}, where the ground-state factor responsible for
the singularity approaching the saddle is treated separately from the
regular excited-state contribution. The difference between such
previous results and our work is that we set up a different asymptotic
expansion, based on large deviations of the canonical contour
integral, which allows us a to achieve a better control on finite-size
corrections in the expansion and to obtain explicit analytical
expressions for the condensate fraction and its fluctuations for a
homogeneous gas in a box.\\


The paper is organized as follows: Sec.~\ref{sec:grand-canonical} will
be devoted to review the standard grand-canonical calculation which is
usually presented in textbooks and to fix the notation which will be
then used throughout the whole paper, identifying the dimensionless
density $\varrho = \lambda_T^3 \rho = \lambda_T^3 N/L^3 $ as the
control parameter for Bose-Einstein condensation in the canonical
framework ($\lambda_T$ is the thermal wavelength); in
Sec.~\ref{sec:canonical-zeta} we will introduce the expression of the
canonical partition function considered for the following calculations
and the definitions of the main observables obtained from it: the
probabilities of the normal and condensed phase, the average number of
bosons in the ground state, i.e., the condensate fraction $\langle
n_0(\varrho) \rangle = \langle N_0(\varrho) \rangle/N$, and the
condensate fraction fluctuations, $\sigma_0(\varrho) = \sqrt{\langle
  N_0^2(\varrho) \rangle - \langle N_0(\varrho) \rangle^2}/N$. In
Sec.~\ref{sec:competing-phases} it will be shown how the explicit
calculation of the normal phase probability, $p_\no(\varrho)$, and of
the condensed phase probability, $p_\co(\varrho)$, can be carried on
respectively in the normal regime, $\varrho<\varrho_c$, using the
saddle-point approximation, and in the condensed regime,
$\varrho>\varrho_c$, using our large-deviations estimate; the same
calculation strategy will be then applied in Sec.~\ref{sec:av-fluct}
to compute the average condensate fraction and of its fluctuations,
deriving analytical expressions with explicit dependence on $\ell = L
/ \lambda_T$, the large-deviations parameter.\\

\section{The grand-canonical calculation}
\label{sec:grand-canonical}

In order to introduce our new asymptotic estimate of the canonical
partition function of the condensate, it is useful to first recall the
main steps of the standard BEC calculation. The starting point of the
calculation is the expression of canonical partition function for a
system of $N$ non-interacting bosons in a three dimensional box of
linear size $L$:

\begin{align}
  \mZ(N) = \sum_{ \substack{ \lbrace n_{\bf k} \rbrace_{{\bf k} \in
        \mathbb{Z}} \\ \sum_{{\bf k}} n_{\bf k} = N }} e^{-\beta
    \sum_{\bf k} n_{\bf k} \epsilon_{\bf k}},
\end{align}
where $\beta=1/(k_BT)$, with $k_B$ the Boltzmann constant and $T$ the
temperature, $\epsilon_{\bf k}=\hbar^2k^2/(2m)$ with $\hbar$ the
reduced Planck constant and $m$ the mass of each bosonic particle, and
$n_{\bf k}$ is the number of the bosons with wave vector ${\bf
  k}$. Here ${\bf k}=(2\pi/L)~{\bf n}$ with ${\bf n}\in \mathbb{Z}^3$.\\

For the purpose of the present calculation it is convenient to write
the gran canonical fugacity $z=e^{\beta\mu}$ in terms of the
dimensionless variable $s=-\beta\mu>0$, i.e., $z=e^{-s}$, which makes
more explicit that the change of ensemble (from canonical to
grand-canonical) amounts to a discrete Laplace Transform, reading as:
\begin{align}
  & \mZ_{\text{GC}}(s)  = \sum_{N=0}^\infty z^{N} \mZ(N) = \sum_{N=0}^\infty e^{-s N} \mZ(N)  \nonumber \\
  & = \sum_{\lbrace n_{\bf k} \rbrace_{{\bf k} \in \mathbb{Z}}} e^{-\sum_{\bf k} n_{\bf k} (\beta\epsilon_{\bf k}+s)} = \prod_{\bf k} \left[ \sum_{n_{\bf k}} \left( e^{-(\beta\epsilon_{\bf k}+s)}\right)^{n_{\bf k}}  \right] \nonumber \\
  & = \prod_{\bf k} \frac{1}{1-e^{-(\beta\epsilon_{\bf k}+s)}},
  \label{eq:discr-Lapl-transf}
\end{align}
so that
\begin{align}
\log \mZ_{\text{GC}}(s) = - \sum_{\bf k} \log\left( 1-e^{-(\beta\epsilon_{\bf k}+s)} \right)
\end{align}
Given the expression of $\mZ_{\text{GC}}(s)$, which can be easily
computed explicitly, one can then retrieve the canonical partition
function by considering the formal definition of the inverse Laplace
transform:
\begin{align}
  \mZ(N) = \frac{1}{2\pi i}\int_{s_0-i\infty}^{s_0+i\infty} ds~e^{s N} \mZ_{\text{GC}}(s),
  \label{eq:inv-Laplace}
\end{align}
where integration is in the complex $s$ plane along the so-called
Bromwich contour, a straight contour parallel to the imaginary axis
and placed to the right of all possible non-analyticities of the
integrand function, which are conventionally assumed to the to the
left of the real value $s_0$. The formula for the inverse Laplace
transform is always formally exact: all the difficulties coming from
the possible lack of equivalence between ensembles, which prevents to
identify the variable $s=-\beta\mu$ with the chemical potential, are
related to the specific method used to compute the contour integral.\\

It is at this stage customary to introduce the grand-canonical
potential
\begin{align}
\Omega(s) = -\frac{1}{\beta}\log \mZ_{\text{GC}}(s),
\end{align}
so to write the inverse Laplace transform as
\begin{align}
  \mZ(N) = \frac{1}{2\pi i}\int_\Gamma ds~e^{H(s)},
  \label{eq:inv-Laplace-Om}
\end{align}
where $\Gamma$ denotes the same Bromwich contour of
Eq.~\eqref{eq:inv-Laplace} and the function $H(s)$ reads as:
\begin{align}
H(s) = L^3 s \rho - \beta \Omega(s),
\end{align}
where $\rho$ is the particle density:
\begin{align}
  \rho = \frac{N}{L^3}.
\end{align}
Let us now recall how the standard saddle-point equation presented in
books is obtained, emphasizing the approximations which can be
improved for a refined large-deviations calculation. In first place
one must isolate the term which, in presence of a vanishing
ground-state energy, can give rise to a non-analyticity. This must be
done before considering the limit where sums are analytically
continued to integrals, thus yielding
\begin{align}
  \Omega(s) &=\frac{1}{\beta} \log\left( 1 - e^{-(\varepsilon_0+s)} \right) + \nonumber \\
  &+ \frac{1}{\beta} \sum_{{\bf k}\neq {\bf 0}} \log\left( 1 - e^{-(\beta\epsilon_{\bf k}+s)} \right),
  \label{eq:grand-potential-2}
\end{align}
where we have introduced the dimensionless energy
\begin{align}
  \varepsilon_0 = \beta \epsilon_{\bf 0}.
  \label{eq:e0-dimless}
\end{align}
The following step is to take continuum limit $L \gg 1$ in the second
term on the right-hand side of Eq.~\eqref{eq:grand-potential-2}, i.e.,
to transform the sum into an integral. By doing this some care must be
taken. That is, a dependence on the scale $L$ must be kept explicitly
into one of the integration extremes in order to carry on
appropriately the large deviations study. By then replacing the sum
with a $3$-dimensional integral in spherical coordinates one gets:
\begin{align}
  &\frac{1}{\beta} \sum_{{\bf k}\neq {\bf 0}} \log\left( 1 - e^{-(\beta\epsilon_{\bf k}+s)} \right) = \nonumber \\
  & \frac{L^3}{4\beta \Gamma(\frac{3}{2}) \pi^{3/2}} \int_{2\pi/L}^\infty dk ~k^{2}~\log\left( 1 - e^{-\left( \beta\frac{\hbar^2 k^2}{2m}+s \right)}\right),
\end{align}
where $\Gamma(x)$ is the Euler Gamma function, i.e.,
$\Gamma(n+1)=n\Gamma(n)$. It is then convenient to introduce the
dimensionless integration variable $r$
\begin{align}
r = k \frac{\lambda_T}{2\sqrt{\pi}},
\end{align}
where $\lambda_T$ is the De-Broglie thermal wavelength:
\begin{align}
\lambda_T = \sqrt{\frac{2\pi\hbar^2}{m k_B T }}
\end{align}
The grand-canonical potential can be then written as
\begin{align}
  & \Omega(s) = \frac{1}{\beta} \log\left( 1 - e^{-(s+\varepsilon_0)} \right) + \nonumber \\
  & \left({L\over \lambda_T}\right)^3 
  \frac{2}{\beta \, \Gamma(\frac{3}{2})} \int_{\sqrt{\pi}\frac{\lambda_T}{L}}^\infty dr ~r^{2}~\log\left( 1 - e^{-r^2-s}\right).
  \label{eq:grand-potential-3}
\end{align}
The expression in Eq.~\eqref{eq:grand-potential-3} suggests that is
more convenient to recast the whole problem in terms of the
dimensionless parameter:
\begin{align}
\ell = \frac{L}{\lambda_T},
\end{align}
and, accordingly, it is convenient to replace the dimensional density
$\rho = N/L^3$ with a dimensionless one
\begin{align}
\varrho = \lambda_T^3  \rho = \lambda_T^3 \frac{N}{L^3} = \frac{N}{\ell^3} , 
\end{align}
so that the canonical partition function can be rewritten as
\begin{align}
  \mZ(\varrho) = \frac{1}{2\pi i}\int_\Gamma ds~e^{\ell^3 h_\ell(s,\varrho)},
  \label{eq:inv-Laplace-Om-adim}
\end{align}
with 
\begin{align}
  h_\ell(s,\varrho) = s \varrho-\frac{1}{\ell^3}\log\left( 1 - e^{-(s+\varepsilon_0)}\right) - q_\ell(s).
  \label{eq:therm-potential}
\end{align}
According to the expression of the thermodynamic potential
$h_\ell(s,\varrho)$, in Eq.~\eqref{eq:inv-Laplace-Om-adim} we wrote
the partition function as a function of the dimensionless density
$\varrho$ rather than of the particle number $N$. In
Eq.~\eqref{eq:therm-potential} we have introduced the function
$q_{\ell}(s)$ defined as:
\begin{align}
q_{\ell}(s) = \frac{2}{\Gamma(\frac{3}{2})} \int_{\sqrt{\pi}/\ell}^\infty dr~r^{2}\log(1-e^{-r^2-s}).
\end{align}

\subsection{BEC as correction to saddle point}

At this step the standard analysis which reveals the presence of
Bose-Einstein condensation proceed as follows.  First one fixes the
desired value of the density $\varrho$ and then considers the
asymptotic value of the potential in the limit
$\ell\rightarrow\infty$, which reads as
\begin{equation}
h_\infty(s) = s \varrho- q_\infty(s).
\end{equation}
The contour integral expression of the canonical partition function is
then computed by means of the saddle-point approximation, solving the
following saddle-point equation
\begin{align}
  \frac{\partial h_\infty(s)}{\partial s} = 0 ~~~\Longrightarrow~~~\varrho = q'_\infty(s^*) = g_{\frac{3}{2}}(e^{-s^*}),
  \label{eq:saddle-point-bec}
\end{align}
where $g_{\frac{3}{2}}(z)$ is the well known Bose function. For all
densities $\varrho$ such that Eq.~\eqref{eq:saddle-point-bec} can be
solved in terms of a real positive value $s^*$, that is in terms of a
real fugacity $z^*=e^{-s^*}$ in the interval $z^* \in [0,1]$, there is
no condensate fraction and the system is perfectly homogeneous in the
thermodynamic limit. The problem is then that for $d=3$ we have
\begin{align}
g_{\frac{3}{2}}(1)=\varrho_c=\zeta(3/2)\approx~2.6124, 
\end{align}
so that no value of the density larger than the critical one,
$\varrho_c$, can be reached in the homogeneous phase. It is this
consideration that led Einstein to the conclusion that a condensate
fraction is needed. In practice one must consider the specific value
of the actual $\ell$ and then, putting first $\varepsilon_0=0$, solve
the following equation
\begin{align}
  \varrho = \varrho_c + \frac{1}{\ell^3}\frac{e^{-s^*}}{1 - e^{-s^*}},
  \label{eq:standard-bec-eq}
\end{align}
which yields the following size-dependent value of the rescaled chemical
potential $s = -\beta \mu$ in the presence of a condensate fraction:
\begin{align}
  s^*_\ell = \frac{1}{\ell^3} \frac{1}{\varrho-\varrho_c}.
  \label{eq:sstar-saddlepoint}
\end{align}
Despite Eq.~\eqref{eq:standard-bec-eq} being presented in most
textbooks as the standard derivation of Bose-Einstein condensation,
this approach is clearly ill-defined in the large-$\ell$ limit. It can
be immediately noticed that any $\ell$-dependent value $s^*_\ell$ of
the chemical potential, fixed from Eq.~\eqref{eq:sstar-saddlepoint} to
obtain a density $\varrho>\varrho_c$ larger than the critical one,
collapses to the same value $s^*=0$ in the thermodynamic limit
$\ell\rightarrow 0$. This is what makes the grand-canonical ensemble
ill defined in the thermodynamic limit.  In what follows we show that
any problem of this approach can be fixed by pursuing a clean analysis
in the canonical ensemble, explaining how the canonical partition
function can be computed in a simple and transparent way for densities
both below, $\varrho<\varrho_c$, and above, $\varrho>\varrho_c$, the
critical one.\\

\section{Canonical Ensemble}
\label{sec:canonical-zeta}

The strategy pursued in this work is to highlight in first place how
the canonical partition function of free bosons can be written as the
sum of two terms representing respectively the partition function of
the {\it normal} phase, $\mZ_{\no}(\varrho)$, and of the condensed
phase, $\mZ_{\co}(\varrho,\varepsilon_0)$. Let us clarify the
distinction between the term {\it phase} and {\it regime} that we will
consider through the whole discussion. With the term {\it normal
  phase} we will always refer to the phase which contains strictly
{\it zero} condensate fraction at all densities, whereas with the term
{\it condensed phase} we will always refer to the phase which, at any
value of the density and at any finite $\ell$, always contains both a
{\it normal fraction}, i.e., bosons in excited states, and a {\it
  condensate fraction}, i.e., bosons in the ground states. In our
system it is then important to distinguish not only the presence of
two of phases but also the presence of two regimes: the normal regime
and the condensed regime. With the term {\it normal regime} we will
indicate the values of the densities below the critical density for
Bose-Einstein condensation, $\varrho < \varrho_c$, and with the term
{\it condensate regime}, we will indicate the values of the density
above the critical one, $\varrho > \varrho_c$. Notice, as will be
clear from the following exposition, that at any finite $\ell$ the
normal regime always contains also the condensed phase, in the same
manner that the condensed regime always contains the normal phase. We
show now that in all regimes the canonical partition function of the
system can be written as:
\begin{align}
  \mZ(\varrho) = \mZ_{\no}(\varrho) + \mZ_{\co}(\varrho,\varepsilon_0).
  \label{eq:Zcan-split}
\end{align}
In Eq.~\eqref{eq:Zcan-split} we kept explicit in the notation the
dependence of $\mZ_{\co}(\varrho,\varepsilon_0)$ on the ground-state
energy $\varepsilon_0$, which will be useful in the following for the
calculation of the average value and the fluctuations of condensate
fraction. The expression of the partition function as the sum of two
terms is obtained by exploiting the trivial mathematical identity
\begin{align}
  \frac{1}{1-e^{-(s+\varepsilon_0)}} = 1 + \frac{e^{-(s+\varepsilon_0)}}{1-e^{-(s+\varepsilon_0)}},
\end{align}
which allows to write 
\begin{align}
  \mZ(\varrho) &= \frac{1}{2\pi i}\int_\Gamma ds~e^{\ell^3 h(s)} = \frac{1}{2\pi i} \int_\Gamma ds~\frac{e^{\ell^3[\varrho s -
        q_\ell(s)]}}{1-e^{-(s+\varepsilon_0)}} \nonumber \\
  & \nonumber \\
  &= \mZ_{\no}(\varrho) + \mZ_{\co}(\varrho,\varepsilon_0)
  \label{eq:Zc-sum}
\end{align}
where
\begin{align}
  \mZ_{\no}(\varrho) &= \frac{1}{2\pi i} \int_\Gamma ds~e^{\ell^3[\varrho s - q_\ell(s)]} \nonumber \\
  \mZ_{\co}(\varrho,\varepsilon_0) &= \frac{1}{2\pi i} \int_\Gamma ds~e^{\ell^3[\varrho s - q_\ell(s)]}~\frac{e^{-(s+\varepsilon_0)}}{1-e^{-(s+\varepsilon_0)}} \nonumber \\
\end{align}
From the above expressions one can compute in all regimes the
probability of the {\it condensed phase} as
\begin{align}
  p_\co(\varrho)= \frac{\mZ_{\co}(\varrho,\varepsilon_0)}{\mZ_{\no}(\varrho)+\mZ_{\co}(\varrho,\varepsilon_0)},
  \label{eq:pcon}
\end{align}
and the probability of the {\it normal phase} as
\begin{align}
  p_\no(\varrho) = \frac{\mZ_{\no}(\varrho)}{\mZ_{\no}(\varrho)+\mZ_{\co}(\varrho,\varepsilon_0)}.
  \label{eq:phom}
\end{align}
The canonical approach makes transparent the first-order features of
Bose-Einstein condensation because it allows to compute the two
probabilities $p_\co(\varrho)$ and $ p_\no(\varrho)$ for all values of
the dimensionless density
\begin{align}
  \varrho = \frac{N}{\ell^3},
\end{align}
showing that they are both finite at any finite $\ell$, both above and
below the critical density $\varrho_c$: this sort of phase coexistence
at finite size both below and above the critical value of the control
parameter is a distinguishing feature of first-order transitions. As we
will show in the next section, the crucial difference between the {\it
  normal} and the {\it condensed} regime is the strategy to compute
$\mZ_{\co}(\varrho,\varepsilon_0)$ and $\mZ_{\no}(\varrho)$. In the
normal regime, $\varrho < \varrho_c$, the partition function and its
derivatives with respect to the ground state energy can be computed in
the large-$\ell$ limit using the saddle-point approximation, allowing
thus to identify the saddle-point parameter $s^* = -\beta \mu^*$ with
the chemical potential of the system. In the condensed regime,
$\varrho > \varrho_c$, it is necessary to consider appropriate
asymptotic estimates of the contour integrals in the complex plane. In
this case the chemical potential cannot be anymore identified with a
saddle-point parameter and the canonical definition $\mu = dF/dN$,
where $F=-T\log\mZ$ is the free energy of the system, must be
considered.  Following for each regime the appropriate calculation
scheme one can then compute the main observables characterizing
Bose-Einstein condensation: the average condensate fraction and its
fluctuations. For the average condensate fraction one has to compute
the average number of bosons in the ground state:
\begin{align}
  \langle N_0 \rangle &= -\frac{\partial}{\partial \varepsilon_0}\log \mZ(\varrho,\varepsilon_0)\bigg|_{\varepsilon_0=0} \nonumber \\
  & = -\frac{1}{\mZ(\varrho)}\frac{\partial \mZ_\co(\varrho,\varepsilon_0)}{\partial \varepsilon_0}\bigg|_{\varepsilon_0=0},
  \label{eq:n0-av}
\end{align}
where an explicit calculation of the derivative in the last line of
Eq.~\eqref{eq:n0-av} yields
\begin{align}
  \frac{\partial \mZ_\co(\varrho,\varepsilon_0)}{\partial \varepsilon_0}\bigg|_{\varepsilon_0=0} = -\frac{1}{2\pi i}\int_\Gamma ds~\frac{e^{\ell^3 [\varrho s - q_\ell(s)]}}{1-e^{-s}} \frac{e^{-s}}{1-e^{-s}},
  \label{eq:n0-av-2}
\end{align}
so that we can even rewrite the expression of the average value as
\begin{align}
\langle N_0 \rangle = -\frac{1}{\mZ(\varrho)}\frac{1}{2\pi i}
  \int_\Gamma ds~e^{\ell^3 [\varrho s - q_\ell(s)]}\frac{e^s}{(e^s-1)^2}.
\end{align}
In the same manner the fluctuations of the condensate fraction
can be computed as
\begin{align}
\langle N_0^2(\varrho) \rangle - \langle N_0(\varrho) \rangle^2 = \frac{\partial^2}{\partial \varepsilon_0^2}\log \mZ(\varrho)\bigg|_{\varepsilon_0=0}\end{align}
where
\begin{align}
  & \frac{\partial^2}{\partial \varepsilon_0^2}\log \mZ(\varrho)\bigg|_{\varepsilon_0=0} = \nonumber \\
  &= \frac{1}{\mZ(\varrho)}\frac{\partial^2 \mZ_\co(\varrho,\varepsilon_0)}{\partial \varepsilon_0^2}\bigg|_{\varepsilon_0=0}
  - \frac{1}{\mZ^2(\varrho)} \left( \frac{\partial \mZ_\co(\varrho,\varepsilon_0)}{\partial \varepsilon_0}\bigg|_{\varepsilon_0=0}\right)^2 
  \nonumber \\
  & = \frac{1}{\mZ(\varrho)}\frac{1}{2\pi i}\int_\Gamma ds~\frac{e^{\ell^3 [\varrho s - q_\ell(s)]}}{1-e^{-s}}
  \frac{e^{-s}(1+e^{-s})}{(1-e^{-s})^2} - \nonumber \\
  & -
  \frac{1}{\mZ^2(\varrho)} \left( \frac{1}{2\pi i}
  \int_\Gamma ds~\frac{e^{\ell^3 [\varrho s - q_\ell(s)]}}{1-e^{-s}} \frac{e^{-s}}{1-e^{-s}}\right)^2. 
  \label{eq:N0-fluct}
\end{align}

In the next section, Sec.~\ref{sec:competing-phases}, will show how to
explicitly compute the two probabilities $p_\no(\varrho)$ and
$p_\co(\varrho)$ in the normal and in the condensed regime. The
subtleties involved in the calculation of the two phases probabilities
will be the same which must be taken into account for the explicit
computation of the condensate fraction $\langle n_0\rangle = \langle
N_0 \rangle/N$ and of its fluctuations $\sigma_0 = \sqrt{\langle N_0^2
  \rangle - \langle N_0 \rangle^2}/N$, which will be then presented in
Sec.~\ref{sec:av-fluct}.\\

\section{Probabilities of phases}
\label{sec:competing-phases}

\subsection{Normal regime}
\label{subsec:cp-normal}

\begin{figure}
  \includegraphics[width=\columnwidth]{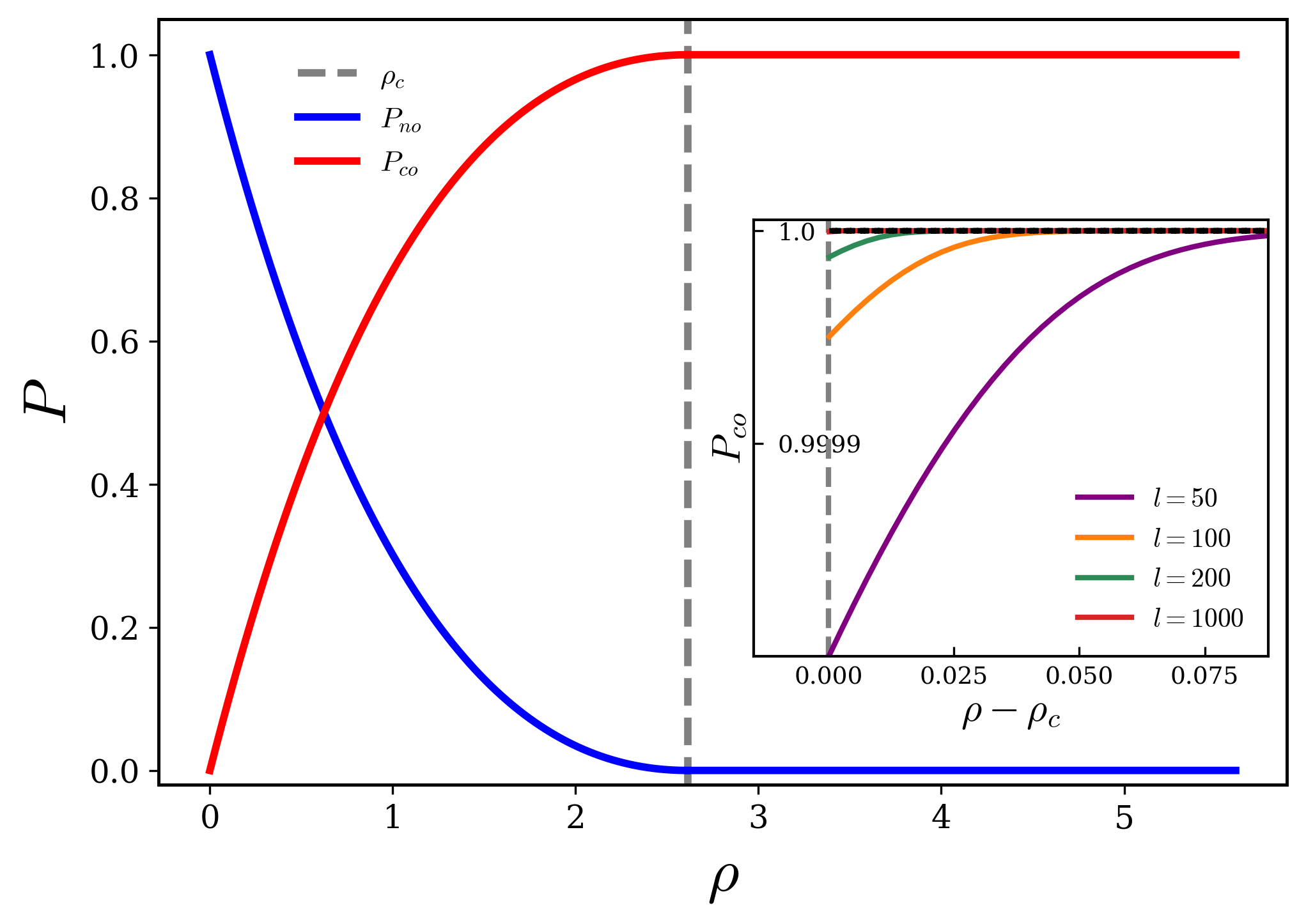}
  \caption{\gcol{{\it Main}: analytical result for probabilities of
      the normal {\it phase}, $p_{\no}(\varrho) =
      \mZ_{\no}(\varrho)/\mZ(\varrho)$, and of the condensed {\it
        phase}, $p_{\co}(\varrho)=\mZ_{\co}(\varrho)/\mZ(\varrho)$,
      for the value $\ell=1000$, where $\mZ_{\no}(\varrho)$ is the
      partition function of the normal phase, $\mZ_{\co}(\varrho)$ is
      the partition function of the condensed phase and $\mZ(\varrho)
      = \mZ_{\no}(\varrho) + \mZ_{\co}(\varrho)$ is the partition
      function of the total system. In the normal {\it regime},
      $\varrho < \varrho_c$, $\mZ_{\no}(\varrho)$ and
      $\mZ_{\co}(\varrho)$ are computed by means of the saddle-point
      approximation, see
      Eqns.~\eqref{eq:Zco-esplicit},\eqref{eq:Zno-esplicit}, while in
      the condensed {\it regime}, $\varrho > \varrho_c$, the two
      partition functions are computed with our large-deviation
      approach for $\ell \gg 1$, see
      Eqns.~\eqref{eq:pno_BEC},\eqref{eq:pco_BEC}. {\it Inset}: Zoom
      in the vicinity of $\varrho_c$ of the parametric dependence on
      $\ell$ of the condensed phase probability $p_{\co}(\varrho)$ in
      the condensed regime, $\varrho > \varrho_c$: $\ell = 50, 100,
      200, 1000$.}}
  \label{fig:probabilities}
\end{figure}

We present in this section a detailed calculation of the partition
functions $\mZ_\no(\varrho)$ and $\mZ_\no(\varrho)$ and of the
corresponding probabilities in the normal regime, i.e., when the
system density is below the critical one, $\varrho < \varrho_c$, and
we can exploit the saddle-point approximation. Such probabilities are
obtained from the expressions of the two corresponding partition
functions, which we recall here for convenience:
\begin{align}
  \mZ_{\co}(\varrho) &= \frac{1}{2\pi i} \int_{s_0-i\infty}^{s_0+i\infty} ds~\frac{e^{\ell^3[s\varrho - q_\ell(s)]}}{e^s-1} \label{eq:Zcon} \\
  \nonumber \\
  \mZ_{\no}(\varrho) &= \frac{1}{2\pi i} \int_{s_0-i\infty}^{s_0+i\infty} ds~e^{\ell^3[s\varrho - q_\ell(s)]}. \label{eq:Zhom}
\end{align}
As anticipated in the previous Sec.~\ref{sec:canonical-zeta}, the two
integrals in Eqns.~\eqref{eq:Zcon},\eqref{eq:Zhom} can be computed in
the normal regime exploiting the saddle-point approximation. For this
reason, let us restore for the expressions in
Eqns.~\eqref{eq:Zcon},\eqref{eq:Zhom} the notation indicating the
Bromwich contour in the following manner:
\begin{align}
\int_\Gamma ds \ldots ~~\rightarrow~~\int_{s_0-i\infty}^{s_0+i\infty} ds \ldots,  
\end{align}
in order to make more explicit the fact that for $\varrho < \varrho_c$
it is always possible to chose a straight integration contour crossing
the real axis at a positive $s_0$ which is the saddle point of the
function. Let us then introduce the two potentials
\begin{align}
  h_{\no}(s,\varrho) & = s\varrho - q_\ell(s) \\
  h_{\co}(s,\varrho) & = s\varrho - q_\ell(s)-\ell^{-3}\log\left( e^s-1\right). 
\end{align}
The two saddle-point equations corresponding to the two integrals in
Eqns.~\eqref{eq:Zcon},\eqref{eq:Zhom}, which allow to determine the
value of the parameter $s$ as a function of density, are therefore
\begin{align}
  \frac{\partial h_{\co}(s,\varrho)}{\partial s}  &= 0 ~\Longrightarrow~s^*_{\co}(\varrho) \nonumber \\
  \frac{\partial h_{\no}(s,\varrho)}{\partial s}  &= 0 ~\Longrightarrow~s^*_{\no}(\varrho), \nonumber 
\end{align}
from which we have, respectively
\begin{align}
 s^*_{\co}(\varrho) ~& \Longleftarrow~\varrho =  q'_\infty(s^*_{\co})+\frac{1}{\ell^3} \frac{e^{s^*_{\co}}}{e^{s^*_{\co}}-1}  \label{eq:sp-co} \\
 s^*_{\no}(\varrho) ~& \Longleftarrow~\varrho =  q'_\infty(s^*_{\no}), \label{eq:sp-ho}
\end{align}
where we have considered the following approximations:
\begin{align}
  q_\ell(s^*) &\approx q_\infty(s^*) = - g_{\frac{5}{2}}(e^{-s^*})\nonumber \\
  q'_\ell(s^*) &\approx q'_\infty(s^*) = +g_{\frac{3}{2}}(e^{-s^*}) \nonumber \\
  q''_\ell(s^*) &\approx q''_\infty(s^*) = -g_{\frac{1}{2}}(e^{-s^*}).
  \label{eq:qell-asymptotics-no}
\end{align}
The saddle-point Eqns.~\eqref{eq:sp-co},\eqref{eq:sp-ho}
can be also written in terms of the fugacity $z=e^{-s}$, reading as
\begin{align}
  \varrho & =  g_{3/2}(z^*_{\co})+ \frac{1}{\ell^3}\frac{1}{1-z^*_{\co}} \label{eq:saddle-fugacity}\\
  \varrho & =  g_{3/2}(z^*_{\no}).
\end{align}
From the two equations above one can immediately read off the
finite-$\ell$ physics of free bosons: the system can be found either
in a normal phase with fugacity $z^*_{\no}$ such that the whole
density is realized by the normal phase, $\varrho =
g_{3/2}(z^*_{\no})$, or in a condensate phase where there is also a
condensate fraction contributing to the total density $\varrho$,
namely
\begin{align}
\varrho = \varrho^{(\no)} + \varrho^{(\co)}
\end{align}
where $\varrho^{(\no)} = g_{3/2}(z^*_{\co})$ and $\varrho^{(\co)} =
\ell^{-3}(1-z^*_{\co})^{-1}$.  The existence of these two phases for
densities $\varrho < \varrho_c$ is clearly a finite-$\ell$ effect.
Indeed from Eq.~\eqref{eq:saddle-fugacity} it is clear that in the
thermodynamic limit the density of the condensate fraction in the
condensate phase vanishes, making this phase perfectly identical to the
normal phase.  What is remarkable is that this simple decomposition of
the partition function as the sum of the condensed and the normal
phase as written in Eq.~\eqref{eq:Zc-sum} not only perfectly describes
the presence of these two phases, but also allows two describe their
probability at any large but finite $\ell$. The probability of the two
phases is obtained from the exact expressions of the corresponding
partition functions, which read respectively as:
\begin{align}
  &\mZ_{\co}(\varrho) = \frac{1}{2\pi i} \left( \frac{2\pi}{-\ell^3 h_{\co}''(s_\co^*)} \right)^{1/2} e^{\ell^3 h_{\co}(s_\co^*,\varrho)} \nonumber \\
  &= - \frac{1}{\sqrt{2\pi \ell^3 h_{\co}''(s_\co^*)}}~e^{\ell^3[s^*_\co\varrho + g_{5/2}(e^{-s^*_\co})]}\frac{e^{-s^*_\co}}{1-e^{-s^*_\co}} \label{eq:Zco-esplicit}\\
  &\nonumber \\
  &\mZ_{\no}(\varrho) = \frac{1}{2\pi i} \left( \frac{2\pi}{-\ell^3 h_{\no}''(s_\no^*)} \right)^{1/2} e^{\ell^3 h_{\no}(s_\no^*,\varrho)} \nonumber \\
  &= - \frac{1}{\sqrt{2\pi \ell^3h_{\no}''(s_\no^*)}}~ e^{\ell^3[s^*_\no\varrho + g_{5/2}(e^{-s^*_\no})]} \label{eq:Zno-esplicit} \\
  & \nonumber 
\end{align}
where
\begin{align}
  h_{\co}''(s_\co^*) & = g_{1/2}(e^{-s_\co^*})+\frac{1}{\ell^3}\frac{e^{s_\co^*}}{(e^{s_\co^*}-1)^2} \label{eq:h2-co}\\
  h_{\no}''(s_\no^*) & = g_{1/2}(e^{-s_\no^*}) \label{eq:h2-no}\\
  & \nonumber 
\end{align}
The probabilities $p_\co(\varrho)$ and $p_\no(\varrho)$ of the two
phases in the normal regime can be finally obtained by first
determining $s^*_\co$ and $s^*_\no$ at the fixed density $\varrho$
from Eqns.~\eqref{eq:sp-co},\eqref{eq:sp-ho} and then plugging the
result into the expressions of
Eqns.~\eqref{eq:Zco-esplicit},\eqref{eq:Zno-esplicit}, thus yielding
\begin{align}
  p_\co(\varrho) &= \frac{\mZ_{\co}(\varrho)}{\mZ_{\co}(\varrho)+\mZ_{\no}(\varrho)} \\
  p_\no(\varrho) &= \frac{\mZ_{\no}(\varrho)}{\mZ_{\co}(\varrho)+\mZ_{\no}(\varrho)}.
  \label{eq:prob-norm}
\end{align}
The behaviour of the two probabilities in Eq.~\eqref{eq:prob-norm} as
a function of $\varrho$, obtained by numerically solving the
saddle-point equation in the normal regime, is show in
Fig.~\ref{fig:probabilities}. In particular, the data for
$p_{\no}(\varrho)$ and $p_{\co}(\varrho)$ as computed from
Eqns.~\eqref{eq:Zco-esplicit},\eqref{eq:Zno-esplicit} are those
plotted for $\varrho < \varrho_c$. \gcol{From the curves shown in the
  main frame of Fig.~\ref{fig:probabilities}, which are computed for
  the value $\ell=10^3$ of the dimensionless large-deviations
  parameter $\ell$, we see how the probability of the condensed phase
  (red increasing line) increases already in the normal regime when
  the dimensionless density $\varrho$ approaches the critical value
  $\varrho_c$ (vertical dotted line), while the probability of the
  normal phase decreases (blue decreasing line). The inset of
  Fig.~\ref{fig:probabilities} shows how for $p_\co(\varrho)$
  finite-size effects are present but small in the condensed
  regime. The formulas for $p_\co(\varrho)$ in the condensed regime
  will be shown in the next section. As we will see, small finite-size
  effects will be found also for the condensate fraction $\langle
  n_0(\varrho) \rangle = \langle N_0(\varrho) \rangle / N$, while they
  will turn out to be crucial for the condensate fraction fluctuations
  of $\sigma_0$ in the condensed regime, $\varrho > \varrho_c$, as
  will be shown at the end of Sec.~\ref{sec:av-fluct}. In what follows
  we show how to compute explicitly $p_\co(\varrho)$ and
  $p_\no(\varrho)$ in the condensed regime.}

\subsection{Condensate regime}
\label{subsec:cp-condensed}

\begin{figure}
    \centering
    \includegraphics[width=\columnwidth]{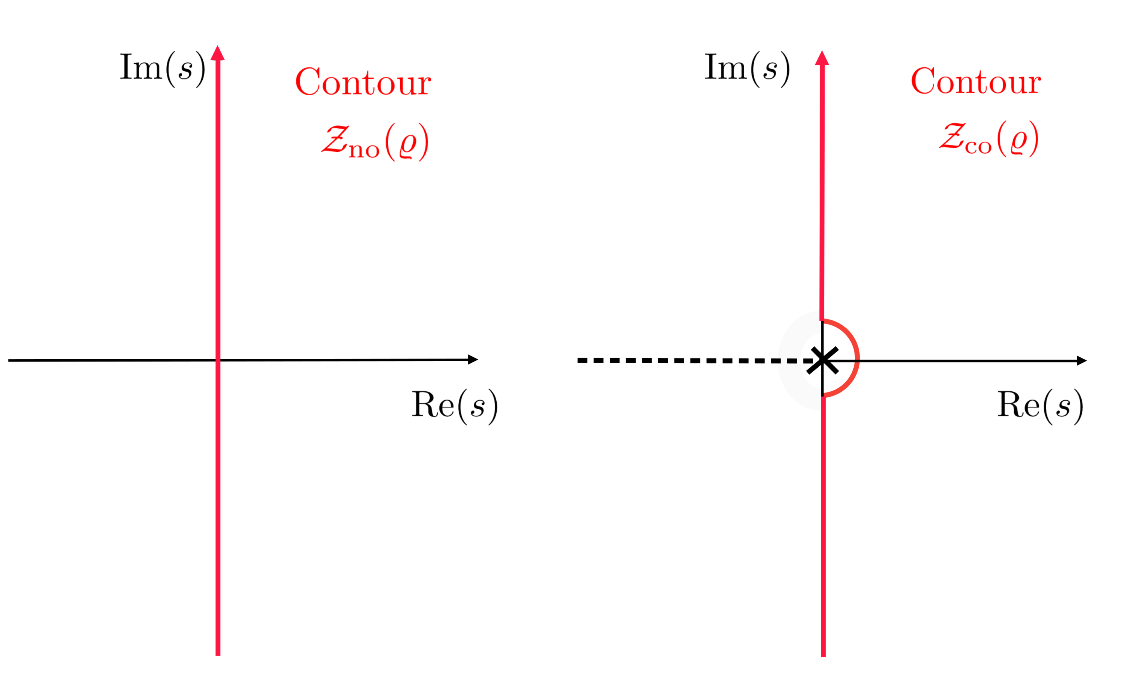}
        \caption{\gcol{In this figure are shown the Bromwich contours
          considered in the condensed regime, $\varrho > \varrho_c$,
          for the explicit calculation of the canonical partition
          function of the normal phase, $\mZ_\no(\varrho)$ (left), and
          for the calculation of the canonical partition function of
          the condensed phase, $\mZ_\co(\varrho)$ (right). The
          splitting of the total partition function in these two
          components, $\mZ(\varrho) = \mZ_\no(\varrho) +
          \mZ_\co(\varrho)$, is always possible and is discussed at the
          beginning of Sec.~\ref{sec:canonical-zeta}. Notice that, for
          $\varrho > \varrho_c$, for the partition function of the
          condensed phase $\mZ_\co(\varrho)$ the contour must be
          wrapped around the branch cut ending at the origin with the
          condensate pole, see the expressions of $\mZ_\co(\varrho)$
          reported in
          Eqns.~\eqref{eq:Zco-recalled},\eqref{eq:Zcon-extended}. Please
          notice that, in agreement with its expression in
          Eq.~\eqref{eq:Zhom}, for $\mZ_\no(\varrho)$ there are no
          singularities on the negative real axis even in the
          condensate phase, so that the integration contour can be
          always taken safely across the origin. All details on the
          complex contour calculations are discussed in
          App.~\ref{app:contours}.}}
    \label{fig:Contour_0}
\end{figure}

The absence of a solution for the equation $\varrho = g_{3/2}(e^{-s})$
for values of the density $\varrho > \varrho_c$ is a signal of the
inequivalence of canonical and grand-canonical ensembles in the
thermodynamic limit, a fact already clearly pointed out in the
historical reference from~Ziff, Uhlenbeck and Kac~\cite{ZUK77}: the
only consistent description of thermodynamics for free bosons at
densities $\varrho > \varrho_c$ is in the fixed density ensemble. In
practice, this means that for the integrals corresponding to the two
partition functions $\mZ_\co(\varrho)$ and $\mZ_\no(\varrho)$ , as
well for those corresponding to the condensate fraction $\langle
n_0(\varrho) \rangle$ and its fluctuations $\sigma_0$, an asymptotic
estimate of the corresponding contour integrals in the complex $s$
plane must be performed explicitly. Let us start the discussion on the
canonical large-deviation approach to the calculation of observables
in the condensed regime from the condensate phase partition function
with zero ground state energy ($\varepsilon_0=0$), $\mZ_\co(\varrho)$,
which we recall here for the ease of the discussion:
\begin{align}
  \mZ_{\co}(\varrho) = \frac{1}{2\pi i} \int_\Gamma ds~e^{\ell^3[\varrho s - q_\ell(s)]}~\frac{e^{-s}}{1-e^{-s}}.
  \label{eq:Zco-recalled}
\end{align}
The overall strategy of the large-deviation calculation is to handle
the integration along the Bromwich contour around the branch cut on
the negative real semi-axis in the complex $s$ plane, see
Fig.~\ref{fig:Contour_0}, by an appropriate small-$s$ expansion of the
integrand function which allows to account for the leading order
non-analiticities in the integration.
The standard strategy of a large-deviation calculation is to restrict
the integration along the Bromwich contour to a neighbourhood of the
origin compatible with deviations of order $\varrho-\varrho_c=\mO(1)$
with respect to $\ell$ for the density. The first step in this
direction consist in expanding the function $q_\ell(s)$ around $s=0$:
\begin{align}
  q_\ell(s) = q_\ell(0) + q_\ell'(0) s + \frac{1}{2} q_\ell''(0) s^2 + \ldots
  \label{eq:qs-taylor}
\end{align}
In first place, if one wishes to account for finite-$\ell$ corrections
to a standard Bose-Einstein calculation, it is necessary to account
for the dependence on $\ell$ for the derivatives at $s=0$ of the
so-called Bose function $q_\ell(s)$. To this aim it is {\it crucial}
to realize that, apart from the order $n=0$ and $n=1$, all derivatives
of the Bose function at $s=0$ correspond to diverging series, if one
consider first the limit $\ell\rightarrow\infty$, since we have
\begin{align}
  q_\infty(s) = - \sum_{k=1}^\infty \frac{e^{-sk}}{k^{\frac{5}{2}}} ~\Longrightarrow~
  q^{(n)}_\infty(0) \propto \sum_{k=1}^\infty k^{n-\frac{5}{2}}
\end{align}
The important observation is that the Bose function $q_\ell(s)$ is
indeed analytic at $s=0$ {\it only} at finite values of the parameter
$\ell$. This observation is crucial: if one, as in the present case,
wishes to perform a large-deviation estimate of the integral in
Eq.~\eqref{eq:Zco-recalled} by expanding $q_\ell(s)$ around the origin
in the complex $s$ plane, it is necessary to first compute the
derivatives $q^{(n)}_\ell(0)$ at $\ell < \infty$ and then account for
the leading order dependence on $\ell$ of the results. All the details
about the explicit calculation of the higher-order derivatives
$q^{(n)}_\ell(0)$ and their dependence on $\ell$ are discussed in
App.~\ref{app:higher-derivatives}. This step is perhaps the most
subtle technical point of the present work: it makes clear that in the
BEC regime, $\varrho > \varrho_c$, beside the non-analyticity due to
the so-called {\it condensate pole}, which is related to the term
$1/(1-e^{-s})$ of the partition function and which does not depend on
$\ell$, one must carefully account also for the non-analiticities of
$q_\ell(s)$ which become manifest in the $\ell\rightarrow\infty$ limit
and must be therefore considered for an accurate large-deviation
estimate of the partition function. Luckily enough, the study of the
leading order dependence in $\ell$ of the higher order derivatives
$q_\ell^{(n)}(s)$, detailed in App.~\ref{app:higher-derivatives},
shows that all derivatives of order $n>2$ are subleading with respect
to the second order term $q_\ell''(0)$ and that the latter grows with
$\ell$ as
\begin{align}
q_\ell''(0)  = - \kappa \ell,  
\end{align}
where $\kappa = 8/\pi$.
On the contrary (as is well known), the first two terms of the
expansion, $q_\ell(0)$ and $q_\ell'(0)$, can be harmlessly replaced
with their thermodynamic limit as done in the normal phase (see
Eq.~\eqref{eq:qell-asymptotics-no}), thus yielding
\begin{align}
  q_\ell(0)~\rightarrow~q_\infty(0) & = -g_{\frac{5}{2}}(1) = -\zeta(5/2)  \\
  q'_\ell(0)~\rightarrow~q'_\infty(0) & = g_{\frac{3}{2}}(1) = \zeta(3/2) 
\end{align}
where $\zeta(z)$ is the Riemann zeta function and the first-order
derivative at $s=0$ corresponds to the critical dimensionless density
for Bose-Einstein condensation in the thermodynamic limit:
\begin{align}
q'_\infty(0)  = \zeta(3/2) = \varrho_c \approx 2.6124.
\end{align}
To the leading order in the large-$\ell$ limit the expansion of
$q_\ell(s)$ can be then written as:
\begin{align}
  q_\ell(s) = -\zeta(5/2) + \varrho_c s - \frac{1}{2}\kappa\ell s^2.
  \label{eq:qs-taylor}
\end{align}
By plugging the expression of $q_\ell(s)$ in Eq.~\eqref{eq:qs-taylor}
into the expression of $\mZ_{\co}(\varrho)$ we get
\begin{align}
  \mZ_{\co}(\varrho) = \frac{e^{\ell^3 \zeta(5/2)}}{2\pi i} \int_\Gamma ds~e^{\ell^3(\varrho -\varrho_c) s + \frac{1}{2}\kappa\ell^4 s^2}~\frac{e^{-s}}{1-e^{-s}}.
  \label{eq:Zcon-extended}
\end{align}
Since we are interested in evaluating the partition function for
deviations above the critical densities of order
\begin{align}
  \varrho -\varrho_c = \mathcal{O}(1),
\end{align}
the strategy of the large-deviation calculation is to consider a
rescaling of the variable $s$ such that
\begin{align}
\ell^3(\varrho -\varrho_c) s \sim 1.
\end{align}
This is obtained by considering the leading term of the integrand
within a neighbourhood of the origin of size $\ell^{-3}$, which can be
done by introducing the following change of variables
\begin{align}
s =  \hat{s}/\ell^3,
\end{align}
and retaining the leading terms of the integration in the large-$\ell$
limit. In terms of $\hs$ the partition function reads
\begin{align}
  \mZ_{\co}(\varrho) = \frac{e^{\ell^3 \zeta(5/2)}}{2\pi i}
  \frac{1}{\ell^3} \int_\Gamma d\hs~e^{(\varrho -\varrho_c)\hs +
    \frac{1}{2}\frac{\kappa}{\ell^2}\hs^2}~
  \frac{e^{-\hs/\ell^3}}{1-e^{-\hs/\ell^3}}.
\end{align}
It is at this stage that the so-called {\it condensate pole} term can
be expanded in series, yielding
\begin{align}
    \frac{e^{-\hs/\ell^3}}{1-e^{-\hs/\ell^3}} =\frac{\ell^3}{\hs}-\frac{1}{2}+\frac{1}{\ell^3}\frac{\hs}{12}-\frac{1}{\ell^9}\frac{\hs^3}{720}+\ldots
\end{align}
By retaining only the first two terms of the above expansion we can
rewrite the partition function as a the sum of two terms, one
containing the condensate pole, now appearing the in form $1/\hat{s}$,
and the other one simply corresponding to a Gaussian integral:
\begin{align}
  \mZ_{\co}(\varrho) &= e^{\ell^3 \zeta(5/2)}~\cpa - \nonumber \\
  &- \frac{e^{\ell^3 \zeta(5/2)}}{2 \ell^3}\frac{1}{2\pi i}\int_{-\i\infty}^{i\infty} d\hs~e^{(\varrho -\varrho_c)\hs + \frac{1}{2}\frac{\kappa}{\ell^2}\hs^2}~
\end{align}
where $\cpa$ is a compact notation for the non-analytic term, for
which the explicit evaluation shown in App.~\ref{app:contours} yields
\begin{align}
  \cpa &= \frac{1}{2\pi i}\int_\Gamma d\hs~\frac{e^{(\varrho -\varrho_c)\hs + \frac{1}{2}\frac{\kappa}{\ell^2}\hs^2}}{\hs} \nonumber \\
  & 1 - \frac{1}{2} \textrm{Erfc}\left( \frac{\ell(\varrho-\varrho_c)}{\sqrt{2\kappa}}\right).
  \label{eq:cp1-text}
\end{align}
Since the Gaussian integration yields
\begin{align}
  \frac{1}{2\pi i}\int_{-i\infty}^{i\infty} e^{(\varrho -\varrho_c)\hs + \frac{1}{2}\frac{\kappa}{\ell^2}\hs^2} =
  \frac{\ell}{\sqrt{2\pi\kappa}} e^{-\ell^2 (\varrho-\varrho_c)^2/(2\kappa)},
\end{align}
the partition function of the condensed {\it phase} in the condensed
{\it regime} can be therefore explicitly written as
\begin{align}
  &\mZ_{\co}(\varrho) = e^{\ell^3 \zeta(5/2)}~\cdot \nonumber \\
  & \cdot~\left[ 1 - \frac{1}{2} \textrm{Erfc}
    \left( \frac{\ell(\varrho-\varrho_c)}{\sqrt{2\kappa}}\right)
  - \frac{1}{2}~\frac{e^{-\ell^2 (\varrho-\varrho_c)^2/(2\kappa)}}{\ell^2\sqrt{2\pi\kappa}} \right] \nonumber\\ 
\end{align}
The partition function of the normal {\it phase} in the condensed {\it
  regime} can be computed following the same strategy for the
expansion, which leads to a simple Gaussian distribution
\begin{align}
  \mZ_{\no}(\varrho) &= \frac{1}{2\pi i} \int_\Gamma ds~e^{\ell^3[\varrho s - q_\ell(s)]} \nonumber \\
  & \cong \frac{e^{\ell^3 \zeta(5/2)}}{2\pi i} \frac{1}{\ell^3} \int_\Gamma d\hs~e^{(\varrho -\varrho_c)\hs + \frac{1}{2}\frac{\kappa}{\ell^2}\hs^2} \nonumber \\
  & = e^{\ell^3 \zeta(5/2)}~\frac{e^{-\ell^2 (\varrho-\varrho_c)^2/(2\kappa)}}{\ell^2\sqrt{2\pi\kappa}}.
\end{align}
From such large-deviations estimates of $\mZ_{\no}(\varrho)$ and
$\mZ_{\co}(\varrho)$ at the leading-$\ell$ order in the condensed
regime we can finally write the explicit expression for the
probability of the normal phase, $p_\no(\varrho)$, and the probability
of the condensate phase, $p_\co(\varrho)$, corresponding to the branch
of the corresponding curves in the condensate regime interval,
$\varrho>\varrho_c$, of Fig.~\ref{fig:probabilities}. Such curves are
the graphical representation of the following explicit formulae:

\begin{align}
  & p_\no(\varrho) = \frac{\frac{e^{-\ell^2 (\varrho-\varrho_c)^2/(2\kappa)}}{\ell^2\sqrt{2\pi\kappa}}}{1 -
    \frac{1}{2} \textrm{Erfc}\left( \frac{\ell(\varrho-\varrho_c)}{\sqrt{2\kappa}}\right) + \frac{1}{2}~\frac{e^{-\ell^2 (\varrho-\varrho_c)^2/(2\kappa)}}{\ell^2\sqrt{2\pi\kappa}}} \label{eq:pno_BEC} \\ \nonumber \\
  & p_\co(\varrho) = \frac{1 - \frac{1}{2} \textrm{Erfc}\left( \frac{\ell(\varrho-\varrho_c)}{\sqrt{2\kappa}}\right)
    - \frac{1}{2}~\frac{e^{-\ell^2 (\varrho-\varrho_c)^2/(2\kappa)}}{\ell^2\sqrt{2\pi\kappa}}}{1 -\frac{1}{2} \textrm{Erfc}\left( \frac{\ell(\varrho-\varrho_c)}{\sqrt{2\kappa}}\right) + \frac{1}{2}~\frac{e^{-\ell^2 (\varrho-\varrho_c)^2/(2\kappa)}}{\ell^2\sqrt{2\pi\kappa}}}. \label{eq:pco_BEC}
\end{align}
The corresponding expression of the canonical partition function of
the whole system in the condensed regime, $\varrho > \varrho_c$, reads
finally as
\begin{align}
  & \mZ(\varrho) = \mZ_\no(\varrho) + \mZ_\co(\varrho) \nonumber \\ &
  = e^{\ell^3 \zeta(5/2)}~\left[ 1 - \frac{1}{2} \textrm{Erfc} \left(
    \frac{\ell(\varrho-\varrho_c)}{\sqrt{2\kappa}}\right) +
    \frac{1}{2}~\frac{e^{-\ell^2
        (\varrho-\varrho_c)^2/(2\kappa)}}{\ell^2\sqrt{2\pi\kappa}}
    \right]
  \label{eq:Ztot-cond}
\end{align}
Let us now discuss the behaviour of the average condensate fraction
and of its fluctuations.

\section{Average condensate fraction and fluctuations}
\label{sec:av-fluct}

\gcol{ In the last, and, in a sense, central section of the present
  work we present explicit formulae for the finite-$\ell$ expressions
  of the average number of particles in the ground state $\langle
  N_0(\varrho) \rangle$ and for their fluctuations $\langle
  N_0^2(\varrho) \rangle - \langle N_0(\varrho) \rangle^2$, both in
  the normal regime, as obtained by means of a standard saddle-point
  approximation, both in the condensed regime, as obtained from the
  large-deviations estimate of the canonical partition function. For
  clarity, we have decided to start this section from a more
  qualitative discussion of the results, with explicit reference to
  figures, leaving for afterwords the detailed derivation of
  formulae.\\

  \subsection{Summary of the results}
  \label{sub:summary}
  
As just said, before going through the detailed derivation of the
analytical formulae we start this section by directly commenting the
two most important figures of this work, Fig.~\ref{fig:n0_average} and
Fig.~\ref{fig:n0_fluctuations}, in which are represented respectively
the average condensate fraction in the canonical ensemble,
\begin{align}
  \langle n_0(\varrho) \rangle  = \frac{\langle N_0(\varrho) \rangle}{N},
\end{align}
and the corresponding fluctuations $\sigma_0(\varrho)$,
\begin{align}
\sigma_0^2(\varrho) = \frac{\langle N_0^2(\varrho) \rangle-\langle N_0(\varrho) \rangle^2}{N^2}.
\end{align}
In the main panel of Fig.~\ref{fig:n0_average} it is shown the
analytical result for the average condensate fraction $\langle
n_0(\varrho) \rangle$, plotted as function of the dimensionless
density $\varrho = N/\ell^3$, for the fixed value of the large
parameter $\ell=10^3$. In the normal regime, $\varrho < \varrho_c$,
the result is obtained with the saddle-point method for large $\ell$,
see Eq.~\ref{eq:n0-avv-reload} in Sec.~\ref{subsec:af-normal}, whereas
in the condensed regime, $\varrho > \varrho_c$, the result is the one
obtained with the large-deviations approach of this work and
corresponds to a plot of the expression in
Eq.~\eqref{eq:N0-complete-cond} of Sec.~\ref{subsec:af-condensate}. In
the main panel of Fig.~\ref{fig:n0_average} the dotted line represents
the asymptotic behavior of $\langle n_0(\varrho) \rangle$ in the
thermodynamic limit $\ell\rightarrow\infty$, as obtained from the
large-$\ell$ leading contribution of the expression in
Eq.~\eqref{eq:N0-complete-cond}, which can be exactly written as
$\langle n_0(\varrho) \rangle = 1 - \varrho_c/\varrho$. What must be
noticed about $\langle n_0(\varrho) \rangle$, which is the condensate
fraction, is that finite-$\ell$ effects for this quantity are quite
small. This is shown in the inset of Fig.~\ref{fig:n0_average}, where
it is represented the parametric dependence on $\ell$ of the
condensate fraction, plotted as a function of $\varrho$ in the
vicinity of $\varrho_c$. In particular, in the inset of
Fig.~\ref{fig:n0_average} it is shown the dependence of the condensate
fraction $\langle n_0(\varrho)\rangle$ on $\varrho$ for the values
$\ell= 50, 100, 200, 1000$. It can be clearly seen that, in the regime
$\varrho > \varrho_c$, the average number $\langle n_0(\varrho)
\rangle$ has very small finite-size effects: already slightly apart
from $\varrho_c$ the curves collapse onto the asymptotic behaviour
$\langle n_0(\varrho) \rangle = 1 - \varrho_c/\varrho$ for all values
of $\ell$.
\begin{figure}
    \centering
    \includegraphics[width=\columnwidth]{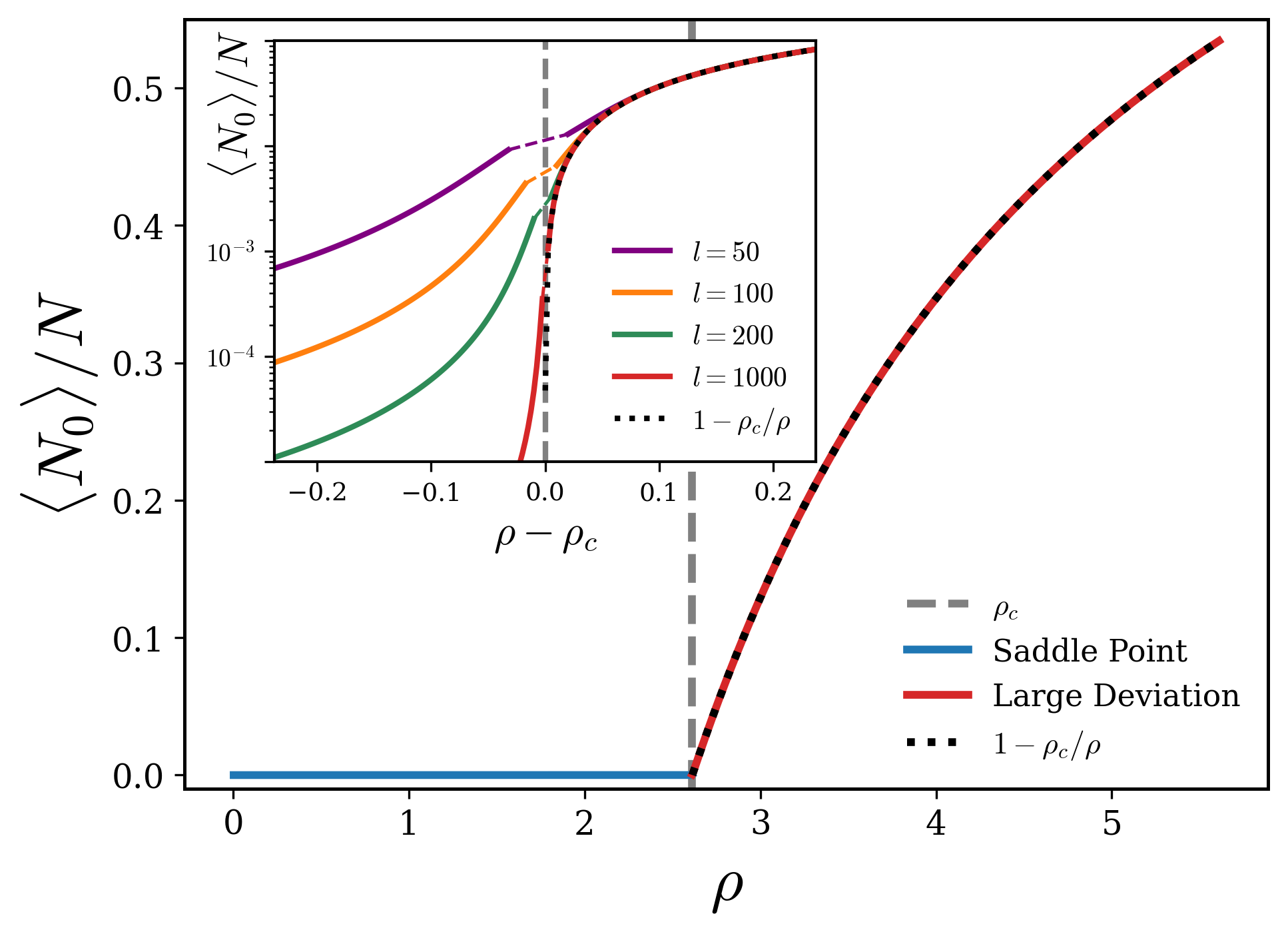}
    \caption{\gcol{{\it Main}: Canonical average of the condensate
        fraction $\langle n_0(\varrho) \rangle = \langle N_0(\varrho)
        \rangle/N$ plotted as function of the dimensionless density
        $\varrho = N/\ell^3$, for the fixed value of the large
        parameter $\ell=10^3$. The blue horizontal continuous line is
        the analytical result for $\langle n_0(\varrho) \rangle$ in
        the normal regime, $\varrho < \varrho_c$, as obtained with the
        saddle-point method for large $\ell$, see
        Eq.~\ref{eq:n0-avv-reload} in Sec.~\ref{subsec:af-normal}. The
        red continuous increasing line is $\langle n_0
        (\varrho)\rangle$ in the condensed regime, $\varrho >
        \varrho_c$, obtained with the large-deviations approach of
        this work and corresponding to the expression in
        Eq.~\eqref{eq:N0-complete-cond} from
        Sec.~\ref{subsec:af-condensate}. Still in the main frame, the
        black dotted line represents the asymptotic behaviour of
        $\langle n_0(\varrho) \rangle = 1 - \varrho_c/\varrho$
        obtained from the analytical expression in
        Eq.~\eqref{eq:N0-complete-cond} in the limit
        $\ell\rightarrow\infty$. {\it Inset}: Zoom around $\varrho_c$
        of the average value $\langle n_0(\varrho) \rangle$ parametric
        dependence on $\ell$, for the values $\ell= 50, 100, 200,
        1000$. It can be clearly seen that, in the regime $\varrho >
        \varrho_c$, the average number $\langle n_0(\varrho) \rangle$
        has very small finite-size effects: slightly apart from
        $\varrho_c$ for all sizes $\ell$ the curves quite well
        superimpose to the asymptotic behaviour $\langle n_0(\varrho)
        \rangle = 1 - \varrho_c/\varrho$.}}
    \label{fig:n0_average}
\end{figure}
The same will not be true for the fluctuations of the condensate
fraction, for which the finite-size effects are crucial, as we are
going to see in a moment. Before commenting the behaviour of
fluctuations let us just briefly recall how our results for the
large-$\ell$ behaviour of the condensate fraction precisely
corresponds to the formula in terms of specific volume reported by
Kerson Huang in the chapter dedicated to Bose-Einstein condensation of
his statistical mechanics book~\cite{H87}:
\begin{align}
  \langle n_0(\varrho) \rangle = 1 - \frac{\varrho_c}{\varrho} = 1 - \frac{v}{v_c},
  \label{eq:n0-fraction-v}
\end{align}
where $v = L^3/N$ is the specific volume. More delicate is, in our
opinion, the rewriting of the same expression in terms of the
temperature $T$. This can be done, at least in principle, by simply
explicitating the dependence of the dimensionless density on the
temperature as
\begin{align}
\varrho(T) = \frac{\rho}{T^{3/2}} \left( \frac{2\pi\hbar}{m k_B} \right)^{3/2}.
\end{align}
The advantage of having done the whole computation in terms of the
dimensionless density $\varrho$ is that such a dependence on $T$ can
be easily explicitated at the end of the calculation. The critical
temperature $T_c$ for Bose-Einstein condensation can be then
identified as the one which, for a fixed particle density $\rho =
N/L^3$, corresponds to the critical {\it dimensionless} density
$\varrho_c$:
\begin{align}
  T_c = \frac{m k_B}{2\pi\hbar}~\left(\frac{\varrho_c}{\rho}\right)^{2/3}.
\end{align}
By doing so, the average condensate fraction can be finally written as
\begin{align}
  \langle n_0(\varrho) \rangle = 1 - \frac{\varrho_c}{\varrho} = 1 - \left(\frac{T}{T_c}\right)^{3/2},
  \label{eq:n0-fraction-T}
\end{align}
which is the standard formula reported in textbooks
as~\cite{H87}. Yet, while Eq.~\eqref{eq:n0-fraction-v} is exact,
Eq.~\eqref{eq:n0-fraction-T} is not. This can be immediately realized
recalling that the first identity of Eq.~\eqref{eq:n0-fraction-T} is
mathematically exact only in the limit $\ell\rightarrow\infty$. But,
since
\begin{align}
\ell = L/\lambda_T \propto T^{1/2} L,
\end{align}
it is then easily argued that the replacement of the dimensionless
densities in the second term of Eq.~\eqref{eq:n0-fraction-T} with the
temperature $T$ in the third term of the same equation is legitimate
only as long as $T^{1/2} L \gg 1$, since the latter is the condition
under which the asymptotic expression is obtained. This tells us that,
as long as a setup with a fixed box size $L$ is considered, there will
be a certain small temperature at which the law of
Eq.~\eqref{eq:n0-fraction-T} will cease to be valid. If one wishes to
obtain asymptotic estimates in the $T\rightarrow 0$ limit for fixed
box sizes, a different kind of expansion should be considered from the
beginning, a point that we are going to discuss
elsewhere~\cite{GLS30}. Luckily, from Fig.~\eqref{fig:n0_average} we
have the evidence that already for not too big values of $\ell$ the
behaviour of the average condensate fraction well superimposes to the
asymptotic law $\langle n_0(\varrho) \rangle = 1 - \varrho_c/\varrho$
at values quite close to $\varrho_c$: this suggest that even if $\ell
\sim T^{2/3}L$ is not too big the asymptotic estimate is a good one.
This means that even in a situation where the box linear size $L$ is
kept fixed, Eq.~\eqref{eq:n0-fraction-T} might yield a good
description of the system down to quite low temperatures.\\

\begin{figure}
    \centering
    \includegraphics[width=\columnwidth]{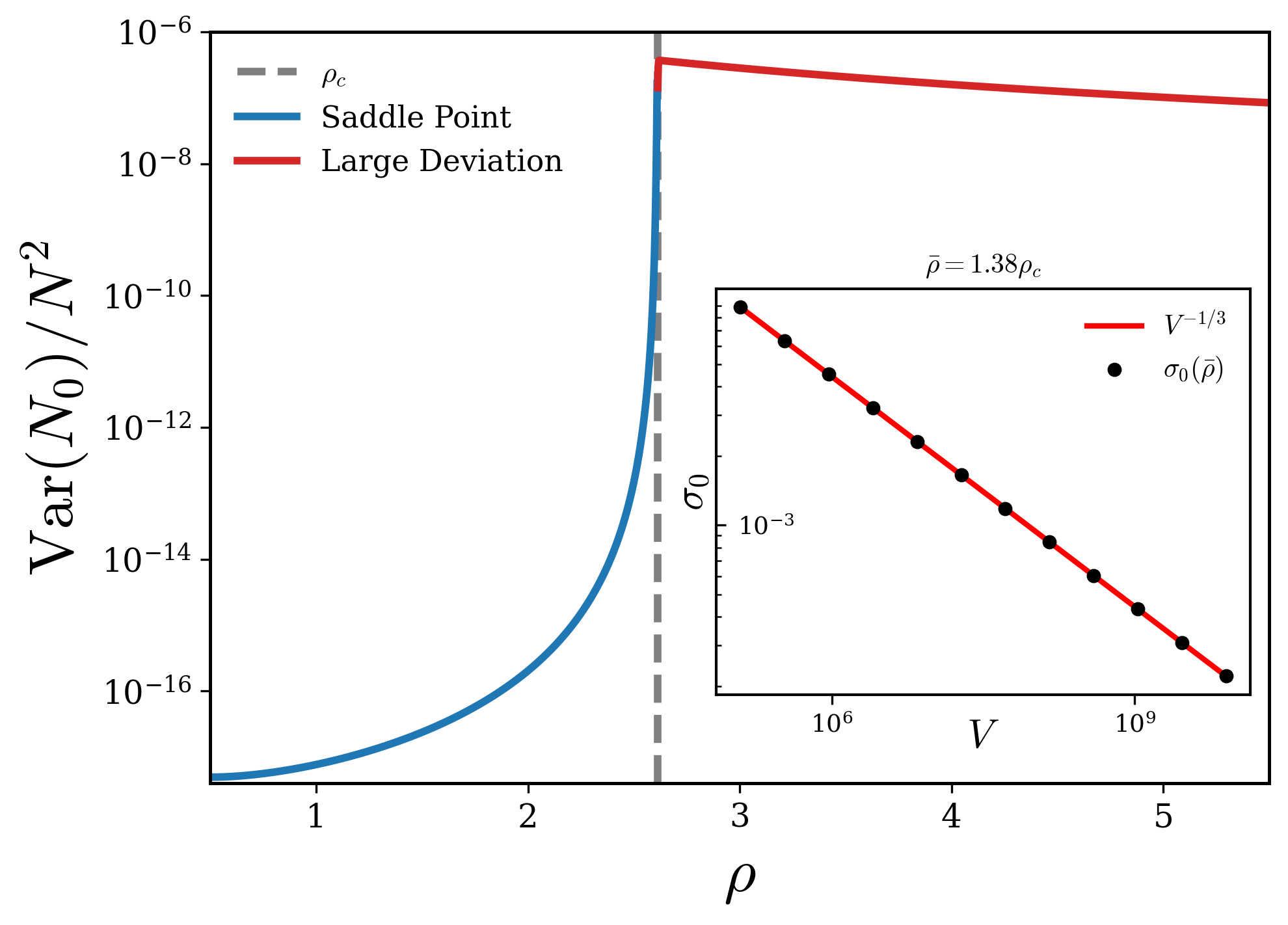}
    \caption{\gcol{ {\it Main}: Canonical average for fluctuations of
        the condensate fraction, $\sigma_0(\varrho) = \sqrt{\langle
          N_0^2 \rangle - \langle N_0\rangle^2}/N$, plotted as
        function of the dimensionless density $\varrho = N/\ell^3$,
        for the fixed value of the large parameter $\ell=10^3$. The
        blue increasing continuous line corresponds to the standard
        saddle-point evaluation of $\sigma_0(\varrho)$ in the normal
        regime, $\varrho < \varrho_c$, see
        Eq.~\eqref{eq:n0-fluct-norm} in
        Sec.~\ref{subsec:af-normal}. The red decreasing continuous
        line corresponds to the large-deviation estimate of
        $\sigma_0(\varrho)$ computed in the condensate regime,
        $\varrho > \varrho_c$, see
        Eqns.~\eqref{eq:N0-complete-cond},\eqref{eq:N02-cond-final} in
        Sec.~\ref{subsec:af-condensate}. The vertical dotted line
        denotes the critical value $\varrho_c$. {\it Inset:} Values of
        $\sigma_0(\varrho)$ are plotted as a function of the volume
        $V=L^3$ for a fixed value of the dimensionless density
        $\tilde{\varrho}=1.38\varrho_c$, in order to show the {\it
          anomalous} scaling $\sigma_0(\varrho) \sim V^{-1/3}$,
        derived in Eq.~\eqref{eq:scaled-fluct-final} in
        Sec.~\ref{sub:summary} of the present paper and consistent
        with the analytical results of~\cite{ZUK77}.}}
    \label{fig:n0_fluctuations}
\end{figure}
Let us now comment the behaviour of the condensate fraction
fluctuations $\sigma_0(\varrho)$: this is the observable for which, in
the condensed regime, the use of the canonical ensemble is crucial. In
the main panel of Fig.~\ref{fig:n0_fluctuations} the continuous blue
increasing curve represents the analytical prediction at $\ell=10^3$
(saddle-point approximation) for $\sigma_0(\varrho)$ in the normal
regime, $\varrho < \varrho_c$, while the continuous red decreasing
branch represents the analytical results (large deviations) in the
condensed regime, $\varrho > \varrho_c$, for the same value of
$\ell$. The inset of Fig.~\ref{fig:n0_fluctuations} shows the {\it
  anomalous} scaling with system size of the condensate fraction
fluctuations at a fixed value of density, $\sigma_0 \sim V^{-1/3} \sim
L^{-1}$, a result consistent with that of Ziff, Kac and
Uhlenbeck~\cite{ZUK77}: as expected for all the experiments where
Bose-Einstein condensation is obtained by reducing the volume at fixed
particle number~\cite{AEMWC95,D9595,BSTH95}, the fluctuations of
condensate fraction vanish in the thermodynamic limit. The fact that
fluctuations are finite at finite $\ell$ while they vanish in the
limit $\ell\rightarrow\infty$ makes their presence entirely a
finite-$\ell$ effect, for which the finite-size analysis of the
problem is crucial. The large-deviations exact estimate detailed in
the following section, Sec.~\ref{subsec:af-condensate}, allows us not
only to recover the correct asymptotic behaviour of the condensate
fraction fluctuations in the large-$\ell$ limit, but to have explicit
formulae with prefactors at large but finite $\ell$, obtained from the
expressions in Eq.~\eqref{eq:N0-complete-cond} and
Eq.~\eqref{eq:N02-complete-cond}, corresponding respectively to
$\langle N_0(\varrho)\rangle$ and $\langle N_0^2(\varrho) \rangle$ and
from which we can get $\sigma_0^2(\varrho) = \langle N_0^2(\varrho)
\rangle - \langle N_0(\varrho) \rangle^2/N^2$. By doing so, we are
able to exactly point out the finite-size correction for a given
system, i.e., to provide precise numbers. Consider for instance a
Bose-Einstein condensate made of Rubidium 87 atoms confined in cubic
box with linear size $L=200~\mu$m at four different temperatures: 10
nK, 50 nK, 100 nK, 200 nK. The smallest temperature, 10 nK, is perhaps
too low to be reached in ordinary experiments on cold atoms, but we
believe it is interesting to report it for illustrational
purposes. From the knowledge of Rubidium mass one can compute the
thermal wavelength $\lambda_T$ for the different temperatures, so that
the value of the large-deviation parameter $\ell = L/\lambda_T$ is
also known and we can use the formulas of
Sec.~\ref{subsec:af-condensate}. In Tab.~\ref{tab:n0-fluctuations} are
reported the values of $\sigma_0(\varrho)$ for the system considered
at the four temperatures just mentioned in the case of two different
values of the dimensionless density: $\varrho_1 = 1.2~\varrho_c$ and
$\varrho_2 = 2~\varrho_c$. The numbers reported in
Tab.~\ref{tab:n0-fluctuations} stress again how, at variance with the
very mild dependence of the average condensate fraction on $\ell$, the
fluctuations $\sigma_0(\varrho)$ are strongly dependent on $\ell$. The
next two subsections will be devoted to a detailed derivation of both
$\langle N_0(\varrho) \rangle$ and $\sigma_0(\varrho)$ in the normal
and in the condensed regime.}\\
\begin{table}
\begin{tabular}{ccccc}
\hline
nK~~&~~$\lambda_T~(\mu\textrm{m})$ ~~& $\ell = L/\lambda_T$ ~&~ $\sigma_0(1.2\varrho_c)$ ~~&~~ $\sigma_0(2 \varrho_c)$ \\
\hline
10 & 1.87 & 107 & $4.7\cdot 10^{-3}$ &~ $2.9\cdot 10^{-3}$ \\
50 & 0.84 & 238 & $2.1\cdot 10^{-3}$ &~ $1.3\cdot 10^{-3}$ \\
100 & 0.59 & 339 & $1.5\cdot 10^{-3}$ &~ $9.1\cdot 10^{-4}$ \\
200 & 0.42 & 476 & $1.1\cdot 10^{-3}$ &~ $6.4\cdot 10^{-4}$ \\
\hline
\end{tabular}
\caption{\gcol{In this Table are reported the values of condensate fraction
  fluctuations $\sigma_0(\varrho) = \sqrt{\langle N_0^2(\varrho)
    \rangle - \langle N_0(\varrho) \rangle^2}/N$ obtained from the
  exact canonical ensemble large-deviations estimates of $\langle
  N_0(\varrho) \rangle$ and $\langle N_0^2(\varrho) \rangle$, whose
  complete analytical expressions are reported respectively in
  Eq.~\eqref{eq:N0-complete-cond} and Eq.~\eqref{eq:N02-cond-final} of
  Sec.~\ref{subsec:af-condensate}. The values of $\sigma_0(\varrho)$
  reported in this Table have been computed for atoms of Rubidium 87
  confined in a box of 200 $\mu$m at four different temperatures and
  for two values of the dimensionless density $\varrho = 1.2\varrho_c,
  2\varrho_c$. Calculations are in done in the approximation of
  periodic boundary conditions for the box.}}
\label{tab:n0-fluctuations}
\end{table}

\gcol{In order to conclude the qualitative discussion of our results
  let us comment on how, from the exact expression of
  $\sigma_0(\varrho)$ obtained in the forthcoming
  Sec.~\ref{subsec:af-condensate}, it is possible to draw the
  asymptotic {\it anomalous} scaling with system volume $V=L^3$ of the
  condensate fraction $\sigma_0(\varrho)$ in the condensed regime. In
  order to do that it is sufficient to retain the leading order terms
  in the large-$\ell$ limit of $\langle N^2_0(\varrho) \rangle$ and
  $\langle N_0(\varrho)\rangle^2$, as computed in
  Eq.~\eqref{eq:N0-complete-cond} and Eq.~\eqref{eq:N02-cond-final},
  which read respectively as:
\begin{align}
  \langle N^2_0(\varrho)\rangle & \approx \ell^6 (\varrho-\varrho_c)^2 + \kappa \ell^4
\label{eq:N02-mean-asymp} \\
 \langle N_0(\varrho)\rangle^2 &\approx \ell^6 (\varrho-\varrho_c)^2,
\label{eq:N0-mean2-asymp}
\end{align}
from which one gets
\begin{align}
  \sigma_0^2(\varrho) &= \frac{\langle N_0^2\rangle - \langle N_0 \rangle^2}{N^2} \approx \frac{1}{\ell^2}~\frac{\kappa}{\varrho} =
  \frac{\kappa}{\lambda_T \rho}\frac{1}{V^{2/3}} \nonumber \\
  \sigma_0(\varrho) &\sim \frac{1}{V^{1/3}}.
  \label{eq:scaled-fluct-final}
\end{align}
The expression in Eq.~\eqref{eq:scaled-fluct-final}, which we have
written both in terms of the dimensionless density $\varrho$ and
dimensionless length $\ell$, both in terms of standard density $\rho$
and the volume $V$, clearly shows us that the condensate fraction
fluctuations $\sigma_0(\varrho)$ in the condensate regime always
vanish for $\varrho > \varrho_c$ in the large volume limit, yet slower
than a central limit theorem would suggest, namely as $V^{-1/3}$
rather than as $V^{-1/2}$. This observation concludes our preliminary
qualitative description of the main results on average condensate
fraction and fluctuations: the two next sections are dedicated to
detailed discussion on the derivation of the results.}

\subsection{Normal Regime: Saddle-Point estimate}
\label{subsec:af-normal}

Let us now start a more detailed exposition of our results from the
study of the average number of particles in the ground state, $\langle
N_0\rangle$, in the normal regime, $\varrho < \varrho_c$. The
expression of $\langle N_0\rangle$ is the one given in
Eq.~\eqref{eq:n0-av}, which for the ease of the reader we recall here
\begin{align}
  \langle N_0(\varrho)\rangle &= \frac{1}{\mZ(\varrho)}\frac{1}{2\pi i}
  \int_\Gamma ds~e^{\ell^3 [\varrho s - q_\ell(s)]}~\frac{e^{-s}}{(1-e^{-s})^2} \nonumber \\
  & = \frac{1}{\mZ(\varrho)}\frac{1}{2\pi i} \int_\Gamma ds~e^{\ell^3 u(\varrho,s)} \nonumber \\
  & = \frac{1}{|\mZ(\varrho)|} \frac{e^{\ell^3 u(\varrho,s_0^*)}}{\sqrt{\ell^3 2\pi u_{\co}''(s_0^*)}},
  \label{eq:n0-avv-reload}
\end{align}
where we have rewritten the whole integrand function in terms of the
potential $u(\varrho,s)$:
\begin{align}
u(\varrho,s) &= \varrho s - q_\infty(s)-\frac{1}{\ell^3}[s+2\log(1-e^{-s})]. \nonumber \\
\end{align}
The saddle-point value $s_0^{*}$ to plug into the last line of
Eq.~\eqref{eq:n0-avv-reload} is therefore the solution of 
\begin{align}
\varrho = g_{\frac{3}{2}}(e^{-s_0^*})+\frac{1}{\ell^3}\frac{1+e^{-s_0^*}}{(1-e^{-s_0^*})},
\end{align}
with the second derivative of the potential appearing in
Eq.~\eqref{eq:n0-avv-reload} reading explicitly as:
\begin{align}
u''(s_0^*) &= g_{\frac{1}{2}}(e^{-s_0^*})+\frac{2}{\ell^3}\frac{e^{-s_0^*}}{(1-e^{-s_0^*})^2}.
\end{align}
The normalization factor $|\mZ(\varrho)|$ in Eq.~\eqref{eq:n0-avv-reload} reads as
\begin{align}
  |\mZ(\varrho)| & = \frac{1}{\sqrt{2\pi \ell^3 h_{\co}''(s_\co^*)}}~e^{\ell^3[s^*_\co\varrho + g_{5/2}(e^{-s^*_\co})]}\frac{e^{-s^*_\co}}{1-e^{-s^*_\co}}
  \nonumber \\
  & + \frac{1}{\sqrt{2\pi \ell^3h_{\no}''(s_\no^*)}}~ e^{\ell^3[s^*_\no\varrho + g_{5/2}(e^{-s^*_\no})]} \label{eq:Z-norm-mod},
\end{align}
where $h_{\no}''(s_\co^*)$ and $h_{\co}''(s_\no^*)$, with the
corresponding solutions $s_\co^*$ and $s_\no^*$, are those written
Eqns.~\eqref{eq:h2-co},\eqref{eq:h2-no} of
Sec.~\ref{subsec:cp-normal}. Similarly, the expression of the
fluctuations $\langle N_0^2(\varrho) \rangle - \langle
N_0(\varrho)\rangle^2$ can be determined by plugging appropriately the
saddle-point solutions into the expression of Eq.~\eqref{eq:N0-fluct},
which yields the final expression:
\begin{widetext}
\begin{align}
   &\langle N_0^2(\varrho) \rangle - \langle N_0(\varrho)\rangle^2 =
  \frac{1}{\mZ(\varrho)}\frac{1}{2\pi i}\int_\Gamma ds~\frac{e^{\ell^3
      [\varrho s - q_\infty(s)]}}{1-e^{-s}}
  \frac{e^{-s}(1+e^{-s})}{(1-e^{-s})^2} - \frac{1}{\mZ^2(\varrho)}
  \left( \frac{1}{2\pi i} \int_\Gamma ds~\frac{e^{\ell^3 [\varrho s -
        q_\infty(s)]}}{1-e^{-s}} \frac{e^{-s}}{1-e^{-s}}\right)^2
  \nonumber \\ & = \frac{1}{|\mZ(\varrho)|}\frac{1}{\sqrt{\ell^3 2\pi
      h_{\co}''(s_\co^*)}}~\frac{e^{\ell^3 [\varrho s^*_\co +
        g_{5/2}(e^{-s^*_\co})]}}{1-e^{-s^*_\co}}
  \frac{e^{-s^*_\co}(1+e^{-s^*_\co})}{(1-e^{-s^*_\co})^2} -
  \frac{1}{\mZ^2(\varrho)}\frac{1}{\ell^3 2\pi h_{\co}''(s_\co^*)}
  \left(\frac{e^{\ell^3 [\varrho s^*_\co +
        g_{5/2}(e^{-s^*_\co})]}}{1-e^{-s^*_\co}}
  \frac{e^{-s^*_\co}}{1-e^{-s^*_\co}}\right)^2, \nonumber \\
  \label{eq:n0-fluct-norm}
\end{align}
\end{widetext}
where a dependence of the saddle-point solutions on the dimensionless
density $\varrho$ is implicitly assumed, i.e.,
$s^*_\co=s^*_\co(\varrho)$ and $s^*_\no=s^*_\no(\varrho)$. The values
of the average condensate fraction $\langle n_0(\varrho) \rangle$ and
of its fluctuations $\sigma_0(\varrho)$ obtained from
Eq.~\eqref{eq:n0-avv-reload} and Eq.~\eqref{eq:n0-fluct-norm} in this
subsection are those plotted for $\varrho < \varrho_c$ respectively in
Fig.~\ref{fig:n0_average} and Fig.~\ref{fig:n0_fluctuations}. Before
moving to the study of average and fluctuations in the condensed
regime one last comment on the behaviour of fluctuations in the normal
regime is in order, in particular how it can be argued from the
expression in Eq.~\eqref{eq:n0-fluct-norm} that they diverge when the
dimensionless density $\varrho$ approaches the critical one from
below, $\varrho \rightarrow \varrho_c$. From the expression of
$|\mZ(\varrho)|$ in Eq.~\eqref{eq:Z-norm-mod} it can be clearly seen
that in the limit $s_{\co}^*\rightarrow 0$, i.e., when $\varrho
\rightarrow \varrho_c$ at fixed $\ell$, the leading order behaviour of
the partition function is
\begin{align}
\varrho\rightarrow\varrho_c ~\Longrightarrow~|\mZ(\varrho)|
~\sim~\frac{1}{s_{\co}^*}.
\label{eq:Zasympt}
\end{align}
From the expression of the fluctuations in
Eq.~\eqref{eq:n0-fluct-norm} and by taking into account the asymptotic
behaviour of $|\mZ(\varrho)|$ for $\varrho$ in the vicinity of
$\varrho_c$ just shown in Eq.~\eqref{eq:Zasympt}, the reader can
easily convince himself that in the limit $s_{\co}^*\rightarrow 0$ the
two terms in Eq.~\eqref{eq:n0-fluct-norm} are diverging in the same
manner:
\begin{align}
\varrho\rightarrow\varrho_c ~\Longrightarrow~\langle N^2_0(\varrho)\rangle \sim \langle N_0(\varrho)\rangle^2 ~\sim ~\frac{1}{(s_{\co}^*)^4},
\end{align}
yet with different prefactors, as is clear from
Eq.~\eqref{eq:n0-fluct-norm}. This guarantees that, approaching the
condensation point $s_{\co}^*=0$ from the normal regime, the whole
expression of the fluctuations diverges as
\begin{align}
  \varrho \rightarrow \varrho_c ~\Longrightarrow~\sigma_0(\varrho) =
  \sqrt{\langle N^2_0(\varrho)\rangle - \langle N_0(\varrho)\rangle^2} ~\sim
  ~\frac{1}{(s_{\co}^*)^4}. \\ \nonumber
\end{align}

\subsection{Condensed Regime: Large-Deviations estimate}
\label{subsec:af-condensate}

The expression of the condensate fraction $\langle
n_0(\varrho)\rangle$ and of its fluctuations $\sigma_0(\varrho)$ in
the condensed regime, $\varrho > \varrho_c$, where the saddle-point
approximation is useless, is finally computed exploiting the same kind
of expansions considered for the study of the canonical partition
function in Sec.~\ref{subsec:cp-condensed}. By expanding the function
$q_\ell(s)$ to second order in $s$ and changing variable to
$\hs=\ell^3 s$, one gets the following expression for the average
number of particles in the ground state $\langle N_0(\varrho)\rangle$:
\begin{align}
  \langle N_0(\varrho) \rangle = \frac{1}{\mZ(\varrho)}\frac{1}{2\pi i}
  \frac{1}{\ell^3}\int_\Gamma d\hs~e^{(\varrho-\varrho_c) \hs + \frac{1}{2}\frac{\kappa}{\ell^2}\hs^2}~
  \frac{e^{-\hs/\ell^3}}{(1-e^{-\hs/\ell^3})^2}\nonumber \\ 
\end{align} 
Expanding then for large $\ell$ to the leading order the condensate
pole term $e^{-\hs/\ell^3}/(1-e^{-\hs/\ell^3})^2$ one gets
\begin{widetext}
\begin{align}
\left\langle N_0(\varrho) \right\rangle &= \frac{1}{\mZ(\varrho)}\frac{1}{2\pi i}
\frac{1}{\ell^3}\int_\Gamma d\hs~e^{(\varrho-\varrho_c) \hs + \frac{1}{2}\frac{\kappa}{\ell^2}\hs^2}
\left( \frac{\ell^6}{s^2} - \frac{1}{12}\right) =
\frac{e^{\ell^3 \zeta(5/2)}}{\mZ(\varrho)} \left[ \ell^3 \cpb - \frac{1}{12} \frac{e^{-\ell^2 (\varrho-\varrho_c)^2/(2\kappa)}}{\ell^2\sqrt{2\pi\kappa}}\right] \nonumber \\
& = \frac{e^{\ell^3 \zeta(5/2)}}{\mZ(\varrho)}\left\lbrace \ell^3  (\varrho-\varrho_c)
  \left[  1 - \frac{1}{2} \text{Erfc}\left( \frac{\ell (\varrho-\varrho_c)}{\sqrt{2\kappa}}\right)\right]
  + \frac{\sqrt{\kappa}}{\sqrt{2\pi}}~\ell^2~e^{- \ell^2(\varrho-\varrho_c)^2/(2\kappa)} - \frac{1}{12} \frac{e^{-\ell^2 (\varrho-\varrho_c)^2/(2\kappa)}}{\ell^2\sqrt{2\pi\kappa}}\right\rbrace \nonumber \\
  &\cong \frac{e^{\ell^3 \zeta(5/2)}}{\mZ(\varrho)}\left\lbrace \ell^3  (\varrho-\varrho_c)
  \left[  1 - \frac{1}{2} \text{Erfc}\left( \frac{\ell (\varrho-\varrho_c)}{\sqrt{2\kappa}}\right)\right]
  + \frac{\sqrt{\kappa}}{\sqrt{2\pi}}~\ell^2~e^{- \ell^2(\varrho-\varrho_c)^2/(2\kappa)} \right\rbrace
  \label{eq:N0-complete-cond}
\end{align}
\end{widetext}
At this stage one is finally left with the evaluation of $\langle
N_0^2(\varrho)\rangle$, which yields

\begin{align}
  & \langle N_0^2(\varrho)\rangle = \frac{e^{\ell^3 \zeta(5/2)}}{\mZ(\varrho)}\frac{1}{2\pi i}\int_\Gamma ds~\frac{e^{\ell^3 [\varrho s - q_\ell(s)]}}{1-e^{-s}} \frac{e^{-s}(1+e^{-s})}{(1-e^{-s})^2}\nonumber \\
  & = \frac{e^{\ell^3 \zeta(5/2)}}{\mZ(\varrho)}\frac{1}{2\pi i}\frac{1}{\ell^3}\int_\Gamma d\hs~\frac{e^{(\varrho-\varrho_c) \hs + \frac{1}{2}\frac{\kappa}{\ell^2}\hs^2}}{1-e^{-\hs/\ell^3}}
  \frac{e^{-\hs/\ell^3}(1+e^{-\hs/\ell^3})}{(1-e^{-\hs/\ell^3})^2}.
\end{align}
In this case the expansion of the condensate pole term yields
\begin{align}
  \frac{e^{-\hs/\ell^3}(1+e^{-\hs/\ell^3})}{(1-e^{-\hs/\ell^3})^3}=
  2 \frac{\ell^9}{\hs^3}-\frac{1}{120}\frac{\hs}{\ell^3}+\ldots
\end{align}
so that, to the leading order in $\ell$, we have the following
expression for $\langle N_0^2(\varrho)\rangle$:
\begin{align}
  \langle N_0^2(\varrho)\rangle & = 2~\frac{\ell^6}{\mZ(\varrho)}\frac{e^{\ell^3 \zeta(5/2)}}{2\pi i}
  \int_\Gamma d\hs~\frac{e^{(\varrho-\varrho_c) \hs + \frac{1}{2}\frac{\kappa}{\ell^2}\hs^2}}{\hs^3} = \nonumber \\ 
  & = 2~e^{\ell^3 \zeta(5/2)} \frac{\ell^6}{\mZ(\varrho)} \cpc,
  \label{eq:N02-cond-final}
\end{align}
where the explicit calculation of the contour integral
reported in App.~\ref{app:contours} yields:
\begin{align}
  \cpc & = \frac{1}{2} \left( (\varrho-\varrho_c)^2 +\frac{\kappa}{\ell^2} \right)
  \left[  1 - \frac{1}{2} \text{Erfc}\left( \frac{\ell (\varrho-\varrho_c)}{\sqrt{2\kappa}}\right)\right] + \nonumber \\
  & + \frac{1}{2} \frac{\sqrt{\kappa}}{\sqrt{2\pi}}~\frac{(\varrho-\varrho_c)}{\ell}~
  e^{- \ell^2(\varrho-\varrho_c)^2/(2\kappa)}.
  \label{eq:N02-complete-cond}
\end{align}
The expression of $\mZ(\varrho)$ appearing at denominator both in the
formula of Eq.~\eqref{eq:N0-complete-cond} for $\langle N_0(\varrho)
\rangle$ and in the formula of Eq.~\eqref{eq:N02-complete-cond} for
$\langle N_0^2(\varrho) \rangle$ is the one written in
Eq.~\eqref{eq:Ztot-cond}.\\

The expression of $\langle N_0(\varrho)\rangle$ and $\langle
N_0^2(\varrho)\rangle$ respectively in Eq.~\eqref{eq:N0-complete-cond}
and Eq.~\eqref{eq:N02-cond-final} can be finally combined together to
yield the expression of the condensate fraction fluctuations
$\sigma_0(\varrho)=\sqrt{\langle N_0^2(\varrho)\rangle- \langle
  N_0(\varrho)\rangle^2}/N$ in the condensed regime, which is the one
plotted in Fig.~\ref{fig:n0_fluctuations} and we have already
thoroughly commented in Sec.~\ref{sub:summary}.\\

\section{Conclusions}
\label{sec:conclusions}

In this work we have shown how the canonical partition function of
free bosons in three dimensions can be exactly computed in the
condensate regime by means of a large deviations asymptotic
estimate. This is the most correct strategy to highlight the nature of
Bose-Einstein condensation as a transition associated to the breakdown
of the equivalence between the fixed density ensemble and the
grand-canonical ensemble, where density is conserved only on
average. The crucial point is to set up the whole large-deviation
approach in terms of the dimensionless parameter $\ell = L
/\lambda_T$, where $L$ is the linear size of the system and
$\lambda_T$ is the thermal wavelength, and of the dimensionless
density $\varrho = N/\ell^3$. In particular, we have highlighted a
specific point which is usually overlooked in all textbook
presentations and in most of the technical literature on this
phenomenon: the non-analyticities of the Bose function $q_\ell(s)$ in
the $\ell\rightarrow\infty$ limit must be appropriately taken into
account to correctly compute the leading-order contribution to the
complex path integration needed to compute the canonical partition
function and the physical observables in the condensate regime, as
thoroughly discussed in Sec.~\ref{subsec:cp-condensed}.  A complete
account of the canonical-ensemble approach to Bose-Einstein
condensation, accompanied by profound observations on the lack of
equivalence between statistical ensembles, was already given in past
within the seminal paper of Ziff, Kac and Uhlenbeck~\cite{ZUK77}: here
we were able to point out the strategy for asymptotic estimates of the
contour integrals which allowed us to reach out analytically for
simpler explicit results. The main advantage of exploiting the
canonical approach to free bosons both in the condensate regime above
the critical density, where it is necessary as is the only consistent
description, both in the normal regime, where the grand-canonical
ensemble is also valid, is to describe the behaviour of the system
within a unique formalism/framework both above and below the critical
density $\varrho_c$. Within this framework the physics of
Bose-Einstein condensation can be explicitly written throughout all
regimes in terms of the competition between two phases, the normal and
the condensed one. The explicit description of Bose-Einstein
condensation in terms of competing probabilities between two phases,
which can only be realized within a canonical ensemble approach, has
the unique advantage of clearly showing the mixed-order character of
this transition. Bose-Einstein condensation is indeed a continuous
transition according to the standard Landau classification, since it
has no latent heat: yet, the present large-deviation approach, which
allows to explicitly cast the problem in terms of competing phases
probabilities, fully reveals the first-order aspects of the
transition. An interesting line of investigation would be a detailed
investigation of how this mechanism is removed by the presence of
interactions, which makes the transition fully second
order~\cite{SP16}. Let us also notice that, in the perspective
provided here, Bose-Einstein condensation bares a strong similarity
with the localization transition in the Discrete Non Linear
Schr\"odinger Equation (DNLSE), which in~\cite{GILM21} has also been
characterized as a mixed-order transition. In both cases the
condensed/localized regime is characterized by the coexistence between
two phases: the homogeneous (DNLSE) or normal phase (BEC) with
extensive entropy/free-energy and the localized (DNLSE) or condensed
(BEC) phase, characterized by subextensive entropy/free energy. In the
case of DNLSE the entropy of the localized phase scales with
$\sqrt{N}$, where $N$ is the number of degrees of freedom,
see~\cite{GILM21}, whereas in the condensed regime of bosons the
free-energy of the condensate fraction vanish in the thermodynamic
limit, as can be easily read of from the expression of
$\mZ_{\co}(\varrho)$ in Eq.~\eqref{eq:Ztot-cond}. It is quite
interesting to notice that the dominance of a phase with subextensive
entropy also characterizes the ergodicity-breaking transition known as
the Random First-Order Transition (RFOT) in models of glasses. The
nature of the RFOT transition is very clearly represented for instance
by the Random Energy Model~\cite{P24}, where the transition can be
typically characterized as a {\it condensation} in phase-space, namely
a phenomenon characterized by the collapse of the statistical measure
on a finite subset of all the microscopic configurations in principle
compatible with the macroscopic constraints. The similarities and
interconnections between these three kind of mixed-order transitions
in different models is surely a matter which deserves further
investigations, as well as the extension of the canonical
large-deviation formalism discussed in the present work to bosons
confined in different kind of geometries, as for instance the standard
harmonic traps often considered in experimental
setups~\cite{AEMWC95,D9595,BSTH95} or the curved geometries which
recently attracted a lot of interest~\cite{TS19}.

\acknowledgements

G.G. and L.S. are partially supported by the ``Iniziativa Specifica
Quantum'' of INFN and by the Project ``Frontiere Quantistiche''
(Dipartimenti di Eccellenza) of the Italian Ministry of University and
Research (MUR). L.S. is also partially supported by the European
Union-Next Generation EU within the National Center for HPC, Big Data
and Quantum Computing (Project No. CN00000013, CN1 Spoke 10: ``Quantum
Computing'') and by the EU Project PASQuanS2 ``Programmable Atomic
Large-Scale Quantum Simulation''.

\clearpage
\appendix

\section{Bose function asymptotics}
\label{app:higher-derivatives} 

This section is dedicated to the calculation of explicit expression
for $q_\ell(s)$ derivatives, in order to show explicitly the leading
order dependence on $\ell$ retained by all derivatives of second and
largest order. In order to be fully pedagogical and connect with
standard calculations shown in all textbooks, let us first compute
explicitly the Bose function in the thermodynamic limit $\ell=\infty$,
which has the following expression:
\begin{align}
q_\infty(s) = \frac{2}{\Gamma(\frac{3}{2})} \int_0^\infty dr~r^{2}\log(1-e^{-r^2-s}).
\end{align}
The strategy for this calculation appearing in all
standard discussion of Bose-Einstein condensation is to consider
the Taylor series of the logarithm for $x \in [0,1]$:
\begin{align}
\log(1-x) = - \sum_{k=1}^\infty \frac{x^k}{k},
\end{align}
thus writing
\begin{align}
  q_\infty(s) &= - \frac{2}{\Gamma(\frac{3}{2})} \int_0^\infty dr~r^{2} \sum_{k=1}^\infty \frac{(e^{-r^2-s})^k}{k} \nonumber \\
  & = -   \sum_{k=1}^\infty \frac{e^{-s k}}{k}
  \left( \frac{2}{\Gamma(\frac{3}{2})}  \int_0^\infty dr~r^{2} e^{-r^2 k} \right) \nonumber \\
  & = - \sum_{k=1}^\infty \frac{e^{-s k}}{k^{\frac{3}{2}+1}} \nonumber \\
  & = - g_{\frac{5}{2}}(e^{-s})
  \label{eq:app-qinfty}
\end{align}
where $g_{\frac{5}{2}}(x)$ is the Bose function of $x$.
While we have that both $q_\infty(0)$ and $q_\infty'(0)$
are finite, corresponding, respectively, to:
\begin{align}
q_\infty(0)  &=  -g_{\frac{5}{2}}(1) = - \zeta(5/2) \approx - 1.3415 \nonumber \\
q_\infty'(s) &=  g_{\frac{3}{2}}(e^{-s}) = \zeta(3/2) = \varrho_c \approx 2.6124
\end{align}
it is then clear from the expression of $q_\infty(s)$ as a sum in
Eq.~\eqref{eq:app-qinfty} that all higher order derivatives at $s=0$
are infinite.\\

Let us now report the expression of $q_\ell(s)$
derivatives at $s=0$ up to fourth order, indicating with
$q^{(n)}_\ell(s)$ the $n$-th derivative.  The objective is to estimate
their dependence on $\ell$ at $s=0$.

In order, we have: 

\begin{align}
q_\ell^{(1)}(s) &= \frac{2}{\Gamma(3/2)} \int_{\sqrt{\pi}/\ell}^\infty dr~r^2~\frac{1}{e^{r^2+s}-1} \nonumber \\ \\
q_\ell^{(2)}(s) &= - \frac{2}{\Gamma(3/2)} \int_{\sqrt{\pi}/\ell}^\infty dr~r^2~\frac{e^{r^2+s}}{(e^{r^2+s}-1)^2} \nonumber \\
q_\ell^{(3)}(s) &= q_\ell^{(2)}(s) + \frac{4}{\Gamma(3/2)} \int_{\sqrt{\pi}/\ell}^\infty dr~r^2~\frac{e^{2 r^2+2 s}}{\left(e^{r^2+s}-1\right)^3} \nonumber \\ 
q_\ell^{(4)}(s) &= q_\ell^{(3)}(s) + \frac{8}{\Gamma(3/2)} \int_{\sqrt{\pi}/\ell}^\infty dr~r^2~\frac{e^{2 r^2+2 s}}{\left(e^{r^2+s}-1\right)^3}\nonumber \\
&- \frac{12}{\Gamma(3/2)} \int_{\sqrt{\pi}/\ell}^\infty dr~r^2~\frac{e^{3 r^2+3 s}}{\left(e^{r^2+s}-1\right)^4} 
\end{align}

From the above we have that the derivatives at $s=0$ simply read as:

\begin{align}
q_\ell^{(1)}(0) &= \frac{2}{\Gamma(3/2)} I_1(\ell) \nonumber \\
q_\ell^{(2)}(0) &= -\frac{2}{\Gamma(3/2)} I_2(\ell) \nonumber \\
q_\ell^{(3)}(0) &= q_\ell^{(2)}(0) +\frac{4}{\Gamma(3/2)} I_3(\ell) \nonumber \\
q_\ell^{(4)}(0) &=  q_\ell^{(3)}(0) + \frac{8}{\Gamma(3/2)} I_3(\ell) -\frac{12}{\Gamma(3/2)} I_4(\ell) 
\end{align}

where we have introduced the function

\begin{align}
I_n(\ell) = \int_{\sqrt{\pi}/\ell}^\infty dr~r^2~\frac{e^{-r^2}}{\left( 1- e^{-r^2}\right)^n}
\end{align}

The asymptotic behaviour for large $\ell$ of the functions $I_n(\ell)$
can be easily determined for all $n>2$ by changing variable and
expanding to first order the exponential in the denominator, yielding

\begin{align}
I_n(\ell ) = c_n\ell^{2n-5}.
\end{align}

The only exception is the case $n=2$, for which, also considering the
importance of this term for the main results of this paper, we show
here how to explicitly compute. By first changing variable we obtain

\begin{align}
I_2(\ell)= \frac{1}{\ell^3} \int_{\sqrt{\pi}}^\infty du~u^{2} \frac{e^{-u^2/\ell^2}}{(1-e^{-u^2/\ell^2})^2}.
\label{eq:second-moment}
\end{align}

The only additional care that we have to pay with respect to the cases
with $n>1$ is that for $n=2$ in order to estimate the dependence of
the integral on $\ell$ we have to expand the exponential at the
denominator up to the second order, thus obtaining:

\begin{align}
  I_2(0) &\simeq  \frac{1}{\ell^3} \int_{\sqrt{\pi}}^\infty du~u^{2}~ \frac{1-u^2/\ell^2+\frac{1}{2}u^4/\ell^4+\ldots}{(u^2/\ell^2-\frac{1}{2}u^4/\ell^4+\ldots)^2} \nonumber \\
  & \simeq \ell \, 2 \int_{\sqrt{\pi}}^\infty {du\over  u^2} =  \ell~\frac{2}{\sqrt{\pi}}.
\end{align}
The purpose of this analysis is to show that the argument of the
exponential function in the expression of $\mZ_\co(\varrho)$ can be
written as
\begin{align}
\ell^3 (s \varrho - q_\ell(s)) = \ell^3 (\varrho - \varrho_c)s + \frac{1}{2}\kappa \ell^4 s^2 + \sum_{n>3}^\infty \alpha_n \ell^{2n-2} s^n.
\end{align}
The above expression, in terms of the rescaled variable $\hs=\ell^3 s$
which one needs for the large deviation estimate of the integral, reads as
\begin{align}
\ell^3 (s \varrho - q_\ell(s)) \rightarrow (\varrho - \varrho_c) \hs + \frac{1}{\ell^2}\left[\frac{\kappa}{2}\hs^2 + \sum_{n>3}^\infty \alpha_n \frac{\hs^n}{\ell^n}\right].
\end{align}
It is clear from the above expression that in the large-$\ell$ limit
all powers of $\hs$ are negligible compared to $\hs^2$, so that they
can all be neglected for a leading order estimate of the integral
corresponding to $\mZ_\co(\varrho)$.

\newpage

\begin{widetext}

\section{Complex contour Integrals} 
\label{app:contours}

\subsection{First Order Pole: $1/s$} 

In this section we recall how to compute the most important integrals
along Bromwich contours which appear in the main text. Let us start
from the condensate pole of order 1, which we call $\cpa$:
\begin{align}
  \cpa &= \frac{1}{2\pi i} \int_\Gamma ds~\frac{e^{(\varrho-\varrho_{c}) s + \frac{\kappa}{2\ell^2}s^2 }}{s} = \nonumber \\
  & = \frac{1}{2\pi i}\left[\int_{i\epsilon}^{i\infty}ds~\frac{e^{(\varrho-\varrho_{c}) s + \frac{\kappa}{2\ell^2} s^2 }}{s} +
    \int_{-i\infty}^{-i\epsilon}ds~\frac{e^{(\varrho-\varrho_{c}) s + \frac{\kappa}{2\ell^2} s^2 }}{s}\right] +
  \frac{1}{2\pi i} \int_{\gamma_\epsilon} ds~\frac{e^{(\varrho-\varrho_{c}) s + \frac{\kappa}{2\ell^2} s^2 }}{s}
\label{eq:Zc-app}
\end{align}
While the first to terms, in the limit $\epsilon\rightarrow 0$
correspond to the principal value, where we have changed variable from
$s$ to $u = e^{-i\frac{\pi}{2}} s$ (rotation counterclockwise of
$\pi/2$ for the contour integration in the complex plane):
\begin{align}
\text{P.V.}\left[\int_{-\infty}^\infty du~\frac{e^{i(\varrho-\varrho_c)u - \frac{1}{2}\frac{\kappa}{\ell^2} u^2}}{u}\right] = i \pi \left(1 - \text{Erfc}\left[ \frac{\ell(\varrho-\varrho_c)}{\sqrt{2\kappa}}\right)\right],
\end{align}
whereas the term in the last line of Eq.~\eqref{eq:Zc-app} yields
simply half (since we have have circle) of the residue at the pole,
namely:
\begin{align}
\frac{1}{2\pi i} \int_{\gamma_\epsilon} ds~\frac{e^{(\varrho-\varrho_{c}) s + \frac{\kappa}{2\ell^2} s^2 }}{s} = \frac{1}{2},
\end{align}
so that we finally have 
\begin{align}
  \cpa &= \frac{1}{2\pi i} \text{P.V.}\left[\int_{-\infty}^\infty du~\frac{e^{i(\varrho-\varrho_c)u - \frac{1}{2}\frac{\kappa}{\ell} u^2}}{u}\right] +
  \frac{1}{2\pi i} \int_{\gamma_\epsilon} ds~\frac{e^{(\varrho-\varrho_{c}) s + \frac{\kappa}{2\ell^2} s^2 }}{s} = \nonumber \\
  &= 1 - \frac{1}{2}\text{Erfc}\left( \frac{\ell (\varrho-\varrho_c)}{\sqrt{2\kappa}}\right)
\end{align}

\subsection{Second Order Pole: $1/s^2$}

The second integral we wish to compute is 
\begin{align}
  \cpb &=\frac{1}{2\pi i} \int_\Gamma ds~\frac{e^{(\varrho-\varrho_{c}) s + \frac{1}{2}\frac{\kappa}{\ell^2} s^2 }}{s^2} 
  \label{eq:cpb-def}
\end{align}
We show here how the calculation of this integral, by means of a simple
integration by parts, can be reconduced to the calculation of $\cpb$.
Indeed, integrating by parts along the complex contour in the complex
plane we have:
\begin{align}
  \cpb = -\frac{1}{2\pi i}\frac{e^{(\varrho-\varrho_{c}) s + \frac{1}{2}\frac{\kappa}{\ell^2} s^2 }}{s}\bigg|_{-i\infty}^{+i\infty} +
  \frac{1}{2\pi i} \int_\Gamma \frac{e^{(\varrho-\varrho_{c}) s + \frac{1}{2}\frac{\kappa}{\ell^2} s^2 }}{s}
  \left[ (\varrho-\varrho_c) + \frac{\kappa}{\ell^2} s \right].
  \label{eq:cpb-1}
\end{align}
Since along the imaginary axis the exponential term decreases as
$e^{-\frac{\kappa}{2}|s|^2/\ell^2}$ the boundary terms vanish for
$s=\pm i\infty$ and one is left with:
\begin{align}
  \cpb & = (\varrho-\varrho_c)~\cpa + \frac{\kappa}{\ell^2}~\frac{1}{2\pi i}
  \int_{-i\infty}^{i\infty} e^{(\varrho-\varrho_{c}) s + \frac{1}{2}\frac{\kappa}{\ell^2} s^2 }  \nonumber \\
  & \nonumber \\
  & = (\varrho-\varrho_c)~\cpa + \frac{\kappa}{\ell} ~\frac{e^{-\ell^2(\varrho-\varrho_c)^2/(2\kappa)}}{\sqrt{2\pi\kappa}}
  \label{eq:cpb-2}
\end{align}
By then plugging the explicit expression of $\cpa$ into Eq.~\eqref{eq:cpb-2} we then get
\begin{align}
  \cpb =  (\varrho-\varrho_c)
  \left[  1 - \frac{1}{2} \text{Erfc}\left( \frac{\ell (\varrho-\varrho_c)}{\sqrt{2\kappa}}\right)\right]
  + \frac{\sqrt{\kappa}}{\sqrt{2\pi}}~\frac{1}{\ell}~e^{- \ell^2(\varrho-\varrho_c)^2/(2\kappa)}.
\end{align}

\subsection{Third Order Pole: $1/s^3$}

We shown here how for the calculation of the contour integral with the
third order pole one can take advantage of the results obtained for
$\cpa$ and $\cpb$ in the previous subsections. Indeed, by integrating
by parts one gets
\begin{align}
  \cpc & =
  \frac{1}{2\pi i} \int_\Gamma ds~\frac{e^{(\varrho-\varrho_{c}) s + \frac{1}{2}\frac{\kappa}{\ell^2} s^2 }}{s^3} = \nonumber\\
  & = -\frac{1}{4\pi i}\frac{e^{(\varrho-\varrho_{c}) s + \frac{1}{2}\frac{\kappa}{\ell^2}s^2 }}{s^2}\bigg|_{-i\infty}^{+i\infty}
  + \frac{1}{4\pi i}\int_\Gamma ds~\frac{e^{(\varrho-\varrho_{c}) s + \frac{1}{2}\frac{\kappa}{\ell^2}s^2 }}{s^2}
  \left[ (\varrho-\varrho_c) + \frac{\kappa}{\ell^2} s \right] = \nonumber \\
  & = \frac{(\varrho-\varrho_c)}{2}~\cpb + \frac{1}{2}\frac{\kappa}{\ell^2}~\cpa 
  \label{eq:cpc-def}
\end{align}
By then plugging the expressions of $\cpa$ and $\cpb$ into Eq.~\eqref{eq:cpc-def}, one gets
\begin{align}
  \cpc &= \frac{1}{2} \left( (\varrho-\varrho_c)^2 +\frac{\kappa}{\ell^2} \right)
  \left[  1 - \frac{1}{2} \text{Erfc}\left( \frac{\ell (\varrho-\varrho_c)}{\sqrt{2\kappa}}\right)\right]
  + \frac{\sqrt{\kappa}}{2\sqrt{2\pi}}~\frac{(\varrho-\varrho_c)}{\ell}~e^{- \ell^2(\varrho-\varrho_c)^2/(2\kappa)}
\end{align}

\end{widetext}

\end{document}